\documentclass[fleqn]{aa}  
\usepackage{graphicx}
\usepackage{txfonts}
\usepackage{natbib}
\usepackage{latexsym}
\usepackage{float}
\usepackage[bookmarks=false]{hyperref}

\newcommand\blfootnote[1]{%
  \begingroup
  \renewcommand\thefootnote{}\footnote{#1}%
  \addtocounter{footnote}{-1}%
  \endgroup
}
\begin{document} 
   \title{Slow magneto-acoustic waves in simulations of a solar plage region carry enough energy to heat the chromosphere}
  \author{
  N. Yadav
          \inst{1, \footnotemark}
          \and
          R. H. Cameron\inst{1}
          \and 
          S. K. Solanki\inst{1,2}
          }
   \institute{
  Max-Planck-Institut f\"ur Sonnensystemforschung,
              Justus-von-Liebig-Weg 3, 37077 G\"ottingen, Germany\\
              \email{yadavn@mps.mpg.de}
         \and
             School of Space Research, Kyung Hee University, Yongin, Gyeonggi 446-701, Republic of Korea\\
             \email{solanki@mps.mpg.de}
             }
   \date{}
  \abstract
{}
   {We study the properties of slow magneto-acoustic waves that are naturally excited due to turbulent convection and
investigate their role in the energy balance of a plage region using three dimensional (3D) radiation-MHD simulations.
   }
   {To follow slow magneto-acoustic waves traveling along the magnetic field lines, we select 25 seed locations inside a strong magnetic element and track the associated magnetic field lines both in space and time. 
   We calculate the longitudinal component (i.e. parallel to the field) of velocity at each grid point along the field line and compute the temporal power spectra at various heights above the mean solar surface.
   Additionally, horizontally averaged (over the whole domain) frequency power spectra for both longitudinal and vertical (i.e. the component perpendicular to the surface) components of velocity are calculated using time-series at fixed locations.
  To compare our results with the observations we degrade the simulation data with Gaussian kernels having FWHM of 100 km and 200 km, and calculate horizontally averaged power spectra for the vertical component of velocity using time-series at fixed locations.
}
   {
  The power spectra of the longitudinal component of velocity, averaged over 25 field lines in the core of a kG magnetic flux concentration, reveal that the dominant period of oscillations shifts from $\sim 6.5$ minutes in the photosphere to $\sim 4$ minutes in the chromosphere. This behavior is consistent with earlier studies restricted to vertically propagating waves. 
  At the same time, the velocity power spectra, averaged horizontally over the whole domain, show that low frequency waves ($\sim 6.5$ minute period) may reach well into the chromosphere. In addition, the power spectra at high frequencies follow a power law with an exponent close to -5/3, suggestive of turbulent excitation.
  Importantly, waves with frequencies above 5 mHz propagating along different field lines are found to be out of phase with each other even within a single magnetic concentration.

  The horizontally averaged power spectra of the vertical component of velocity at various effective resolutions show that the observed acoustic wave energy fluxes are underestimated, by a factor of three even if determined from observations carried out at a high spatial resolution of 200 km. 
  Since the waves propagate along the non-vertical field lines, measuring the velocity component along the line-of-sight, rather than along the field contributes significantly to this underestimate.
  Moreover, this underestimation of the energy flux indirectly indicates the importance of high-frequency waves that are shown to have a smaller spatial coherence and are thus more strongly influenced by the spatial averaging effect compared to low-frequency waves. 
  }
   {Inside a plage region, on average a significant fraction of low frequency waves leak into the chromosphere due to inclined magnetic field lines.
Our results show that longitudinal waves carry (just) sufficient energy to heat the chromosphere in solar plage. However, phase differences between waves traveling along different field lines within a single magnetic concentration can lead to underestimates of the wave energy flux due to averaging effects in degraded simulation data and similarly, in observations with lower spatial resolution. 
We conjecture that current observations (with spatial resolution around 200 km) underestimate the energy flux by roughly a factor of three, or more if the observations have lower spatial resolution.
   We expect that even at very high resolution, such as expected with the next generation of telescopes, such as DKIST and the EST, on average less than half of the energy flux carried by such waves will be detected if only the line-of-sight component of the velocity is measured. 
   }
   \keywords{Sun: chromosphere -- Sun: faculae, plages -- Sun: magnetic fields -- Methods: numerical.}
 \titlerunning{Slow magneto-acoustic waves in the plage chromosphere}
   \maketitle
   
\section{Introduction}
\blfootnote{ \textbf{$^\ast$} Current address: Centre for mathematical Plasma Astrophysics, Department of Mathematics, KU Leuven, Celestijnenlaan 200B, B-3001 Leuven, Belgium}
The solar surface is populated with magnetic structures existing at various spatial scales e.g., large sunspots, intermediate sized pores and plage magnetic features, and small-scale internetwork magnetic bright points etc.
The magnetic field concentrated in such features plays an important role in the wave energy transport from the turbulent convection zone to the upper layers of the solar atmosphere (\citealt{solanki2006,2013Mathioudakis}).
Beneath the solar surface, acoustic p-modes are generated by turbulent convection (\citealt{1977ApJGoldreich,2001Nordlund}).
At the layer where the sound speed is equal to the Alfv\'en velocity, these incoming acoustic waves or p-modes are partly ``converted'' into fast waves of a magnetic nature and partly ``transmitted'' as slow waves without changing their acoustic nature (\citealt{cally2007, khomenko2012}). 
Propagating upward through the solar atmosphere, the  amplitude  of acoustic  waves  increases    because  of  the  rapid  decrease in  density  with  increasing  height and are dissipated by shock formation (\citealt{2009vecchio}).

Decades ago, \citet{biermann1948} and \citet{Schwarzschild1948} suggested this acoustic wave-based heating mechanism as a plausible means of heating the non-magnetic chromosphere of the quiet Sun.
Their proposal was followed by several numerical and observational studies (\citealt{Ulmschneider2005}; \citealt{Praderie1976}) and was long considered an important heating source in the lower chromosphere.
Nonetheless, using the observations from the Transition Region and Coronal Explorer (TRACE; \citealt{Handy1999}), \citet{fossumnature2005, fossum2005,fossum2006} concluded that the wave energy flux of high-frequency acoustic waves is insufficient to account for the radiative losses in the non-magnetic solar chromosphere and hence cannot constitute the dominant heating mechanism.
Later, \citet{cuntz2007} criticized these results and suggested that high-frequency acoustic waves can indeed be sufficient to heat the  nonmagnetic  solar  chromosphere. 

With increasing observational capabilities, interest in investigating the role of high-frequency (f > 10 mHz) magneto-acoustic waves in chromospheric heating was revived.
Using observations from the G\"ottingen Fabry-Perot spectrometer (FPI) at the German Vacuum Tower Telescope, \citet{gonza2009,gonza2010a} found significant energy flux ($\sim$3000 $W/m^2$) in frequencies above the acoustic cutoff frequency ($\sim$5.2 mHz).
Then, analyzing the observations recorded by the Imaging Magnetograph eXperiment (IMaX) spectropolarimeter onboard the {\sc Sunrise} balloon-based observatory (\citealt{sunrise2010,Bathol2011,Gandorfer2011,Berkefeld2011}), \citet{gonza2010b} strengthened the idea that acoustic waves with periods shorter than the acoustic cutoff-period can significantly contribute to the heating of the solar chromosphere.
However, they could not conclusively verify that the acoustic wave energy flux is sufficient to compensate for the radiative losses of the non-magnetic quiet solar chromosphere.

\begin{figure*}
  \centering
        \includegraphics[scale=0.5]{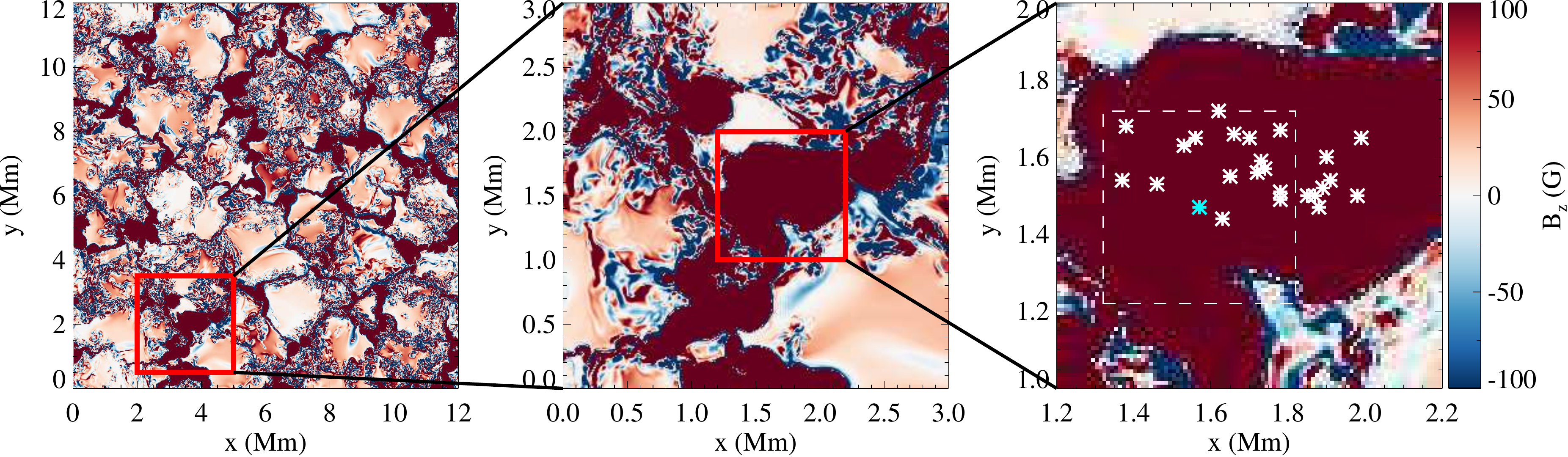}
      \caption{Left: 2D map of the vertical component of the magnetic field vector at $z$=0 layer (corresponding to average $\tau=1$ layer) covering the whole horizontal extent of the simulation domain (saturated at $\mathrm{\pm}$ 100 G), Middle: Blowup displaying the sub-region (red box in the left panel) selected for magneto-acoustic wave analysis. Right: Blowup displaying the seed locations (see main text for details). The dashed square in the rightmost panel covers the region shown in Fig. \ref{coherence}}
         \label{bz_maps}
  \end{figure*}
 \begin{figure}
     \centering
     \includegraphics[scale=0.16]{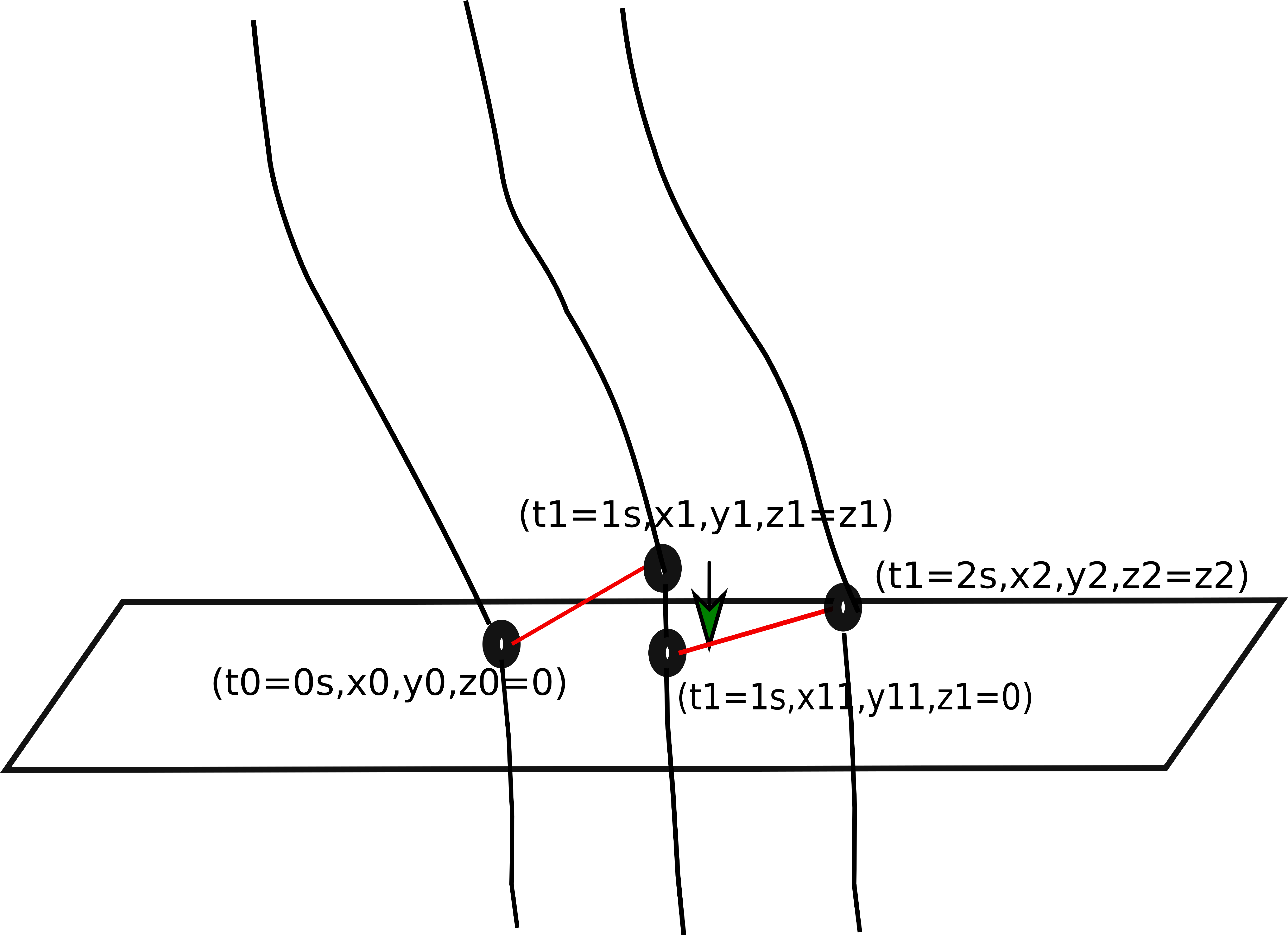}\\
     \caption{Cartoon displaying the scheme used to track a single magnetic field line over time}
     \label{cartoon}
 \end{figure}
\begin{figure*}
    \centering
    \includegraphics[scale=0.11]{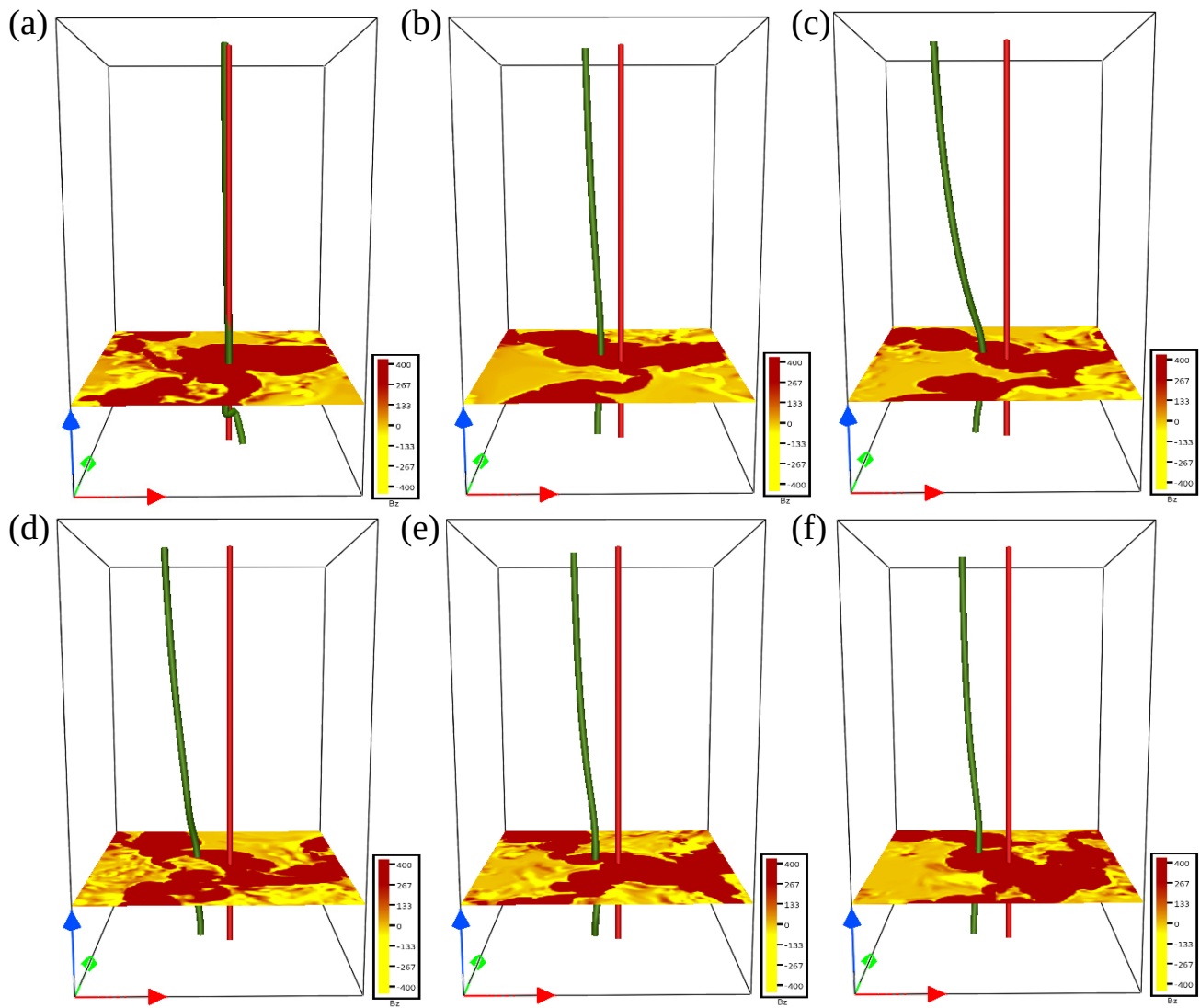}
   \caption{Three-dimensional visualization of magnetic field line (green) corresponding to the seed point shown as a cyan asterisk in Fig. \ref{bz_maps}.
   The horizontal extent corresponds to the region shown in Fig. \ref{bz_maps}b. Panels (a-f) correspond to snapshots at every 5 minutes starting from t=0 (the same instance as in Fig. \ref{bz_maps}). The red vertical line corresponds to the initial location of the seed point. Saturated maps of magnetic field strength at the mean solar surface are also shown for each temporal snapshot.}
   \label{vapor}
\end{figure*}
 More recently, using high spatial and temporal resolution data from the {\sc Sunrise} Filter Imager (SuFI, \citealt{Gandorfer2011}), \citet{shahin2017} investigated the propagation of acoustic waves in the lower solar atmosphere by analyzing both horizontal-displacement oscillations of magnetic bright points and their intensity perturbations.
They observed upward propagating compressible and incompressible (kink) waves at high frequencies (up to 30 mHz).
Previously, using numerical modeling, \citet{hasan2005} had proposed the existence of short-period magneto-acoustic waves ($\sim$ 20s) and suggested that they are excited due to the random displacements of magnetic field lines caused by intergranular turbulence.
\citet{Lawrence2011} analyzed the data recorded with the Rapid Oscillations in the Solar Atmosphere (ROSA) instrument and supported the idea of high-frequency acoustic wave excitation due to photospheric turbulence and showed that spectral power in the 28-326 mHz range follows a power law with exponent $\mathrm{-1.21\pm0.02}$, consistent with the presence of turbulent motions.
Overall, there have been numerous observational and simulation studies over the last half-century to quantify the role of slow magneto-acoustic waves in the heating of solar chromosphere.
However, possibly due to the limited spatial resolution and cadence from which observations suffer, these waves have never been found to carry sufficient energy flux to explain chromospheric heating. 

Most of the aforementioned wave studies correspond to quiet Sun and magnetic bright points, however, acoustic waves play an important role in heating the magnetic chromosphere as well (\citealt{1993Lites,Kalkofen2007}). 
The active region chromosphere is filled with small-scale, short-lived jet-like features called fibrils (e.g., \citealt{Pietarila}) that are suggested to make a non-negligible contribution to the heating of the solar chromosphere (\citealt{1983withbroe, depontieu2004}).
These fibrils (spicules when located on the solar limb) are driven by magneto-acoustic shocks formed by leakage of upward propagating slow magneto-acoustic waves from the photosphere (\citealt{viggo2006, depontieu_2007}).

Moreover, acoustic oscillations in the photosphere do not always display the same periods as in the chromosphere.
For example, 5-minute oscillations are not observed in the chromosphere of sunspots (\citealt{1972Bhatnagar, 2000bogdan, bogdan2006} and references therein). 
\citet{1986lites} showed that regions with high oscillatory power in the 3-min band are uncorrelated with the regions of high oscillatory power in the 5-min band within the umbrae, which suggests that the 5-min oscillations in the photosphere do not excite the 3-min oscillations in the chromosphere of sunspots. 
However, the high-frequency peaks of the photospheric power spectra are often correlated with the peaks in the power spectra of chromospheric oscillations (\citealt{1985lites,2006Centeno}).
\citet{2006Centeno} performed phase diagnostics of the chromospheric and the photospheric oscillations and suggested that the 3-min oscillations observed in the chromosphere come from the photosphere by means of linear wave propagation, rather than from nonlinear interaction of 5-min modes.
Using time-series observations of various spectral lines originating at different heights in the solar atmosphere, \citet{centeno2009} investigated the propagation of magneto-acoustic waves in different magnetic regions on the Sun.
They showed that photospheric oscillations have similar characteristics in all the regions, whereas, chromosopheric oscillations depend on their magnetic properties (e.g. strength and inclination of magnetic field through which they propagate).
Power spectra of chromospheric oscillations in the plage regions are shown to exhibit a dominant peak in the 5-minute band in contrast to sunspots and quiet Sun (\citealt{1972Bhatnagar,centeno2009}).
\citet{1993Lites} analyzed and compared the solar oscillations between the chromospheric network and internetwork regions using 1 hr sequence of spectrograms of a quiet Sun region.
They found internetwork chromospheric velocity power spectrum is dominated by high-frequency ($\sim$ 5 mHz) oscillations while power spectra in network regions are dominated by low-frequency oscillations with periods of 5-20 min. Similar results are recently obtained by \citet{2019rajaguru} using Helioseismic and Magnetic Imager and lower-atmospheric UV emission maps in the 170 and 160 nm channels of the Atmospheric Imaging Assembly, both onboard the Solar Dynamics Observatory of NASA.

The shift of the dominant wave period from 5-minute in the photospheric oscillations to 3-minute in the chromospheric oscillations is often attributed to the cut-off period of the temperature minimum (\citealt{1973SMichalitsanos,1977bel,fleck1991,verth_ch25,2006mcintosh,Khomenko2015,Chae_2017}).
The effective cut-off period may, however, increase due to inclination of the magnetic field relative to the vertical (\citealp{Schwartz_1984,depontieu2004,depontieu2005}) or due to departure from adiabaticity (\citealt{khomenko2008,2020Felipe}), resulting in the propagation of waves with periods of more than 3-minutes.
\citet{2010wijn} showed that waves with three-minute periods propagate in the central facular chromosphere where magnetic field is more vertical while five-minute waves propagate at the peripheral regions where the magnetic field is both inclined and expanding.

In this paper, we investigate the properties of slow magneto-acoustic waves in radiation-MHD simulations of a plage region by tracking magnetic field lines in both space and time.
We compare the temporal power spectra of longitudinal velocity averaged over 25 field lines inside a strong magnetic element with the horizontally averaged (over the whole domain) power spectra.
To compare with the observations, we degrade our simulation data and calculate the power spectra of the vertical component of velocity and compare it with power spectra obtained with full resolution data.
In addition, we estimate the acoustic energy flux associated with slow magneto-acoustic waves for full resolution as well as for the degraded data.

This paper is organized as follows: Sect. \ref{2} comprises the details of numerical simulations and methodology used for field line tracking, results are presented and discussed in Sect. \ref{3}, and summary of the work done and concluding remarks are given in Sect. \ref{4}.

\section{Methodology}\label{2}
\subsection{Simulation Setup}
We have used the 3D radiation-MHD code MURaM  (\citealp{voegler2005, Rempel2014, Rempel2017}) for simulating a unipolar plage region.
The basic ingredients of the MURaM code are the MHD equations that are solved using a $4^{th}$ order finite difference scheme along with a short characteristic radiative transfer scheme.
This code takes into account most of the essential physics required to simulate the upper convection zone where acoustic waves are excited and the lower solar atmosphere where magneto-acoustic waves transform into shocks.
The simulation domain is 12 Mm $\times$ 12 Mm $\times$  4 Mm (with the third dimension referring to the vertical z-direction) with the top boundary located at 2500 km above the average $\tau=1$ layer (continuum formation height). 
The simulation domain is resolved by 1200 $\times$ 1200 $\times$ 400 grid points with a grid spacing of 10 km in all three directions.
The simulation domain is periodic in the horizontal directions with an open bottom boundary and closed top boundary.
For the magnetic field, we have used a vertical magnetic field as the top boundary condition.

To simulate a unipolar plage region, we ran the hydrodynamic simulations for almost 2 hrs to reach a statistically relaxed state and then introduced a uniform vertical magnetic field of 200 G and ran it again for 1.2 hrs, so that the system again reaches a statistically stationary state.
Then, to investigate wave properties, we collect a sequence of 25 minutes at 1 s cadence for a sub-region of 3 Mm $\times$ 3 Mm $\times$ 3 Mm that extends from 500 km below the mean solar surface to 2.5 Mm above the mean solar surface.
We chose a smaller sub-region due to disk space and memory constraints but did not compromise on temporal cadence and spatial resolution for the purpose of capturing high-frequency waves.
Figure \ref{bz_maps} (left panel) displays a map of the vertical component of the magnetic field at the mean solar surface (z=0) at the beginning of the sequence.
The selected sub-region chosen for the wave study corresponds to the red box in the left panel and is shown on an enlarged scale in the middle panel of Figure \ref{bz_maps}.
The right panel of this figure displays the region where we have placed the seed points (shown as asterisks) to perform the wave analysis (described in the next subsection). 

\subsection{Wave analysis: Field line tracking}\label{wave analysis}
The solar photosphere is highly turbulent and dynamic. 
Granular buffeting and turbulent plasma motions excite a wide variety of waves over a large frequency range that may dissipate and heat the solar atmosphere.
Many of the previous numerical simulations invoke wave excitation sources e.g., transverse and torsional drivers mimicking the granular buffeting and photospheric vortex motions, respectively (\citealp{vigeesh2012}).
These drivers perturb the photospheric foot-points of magnetic flux tubes, exciting perturbations within them that then travel to higher layers along the magnetic field lines in the form of MHD waves.
However, there can be many other sources that may excite MHD waves viz. granular buffeting, intergranular turbulence (\citealt{hasan2000,chitta_2012}), intermittent impulsive events (\citealt{shahin2013}), magnetic pumping (\citealp{kato2011,kato2016}), mode conversion and transmission (\citealp{cally2007,khomenko2012}), the interaction of the magnetic elements with the convective flows (\citealt{ballegoijen2011}), etc.
Therefore, it is crucial to include these additional wave-excitation sources in numerical models in as natural a way as possible, i.e. without having to prescribe them. If they are included roughly as they occur on the Sun, we would be able to quantify the net energy flux being carried by these waves and their importance in heating the chromosphere.

\begin{figure}
     \centering
     \includegraphics[scale=0.115]{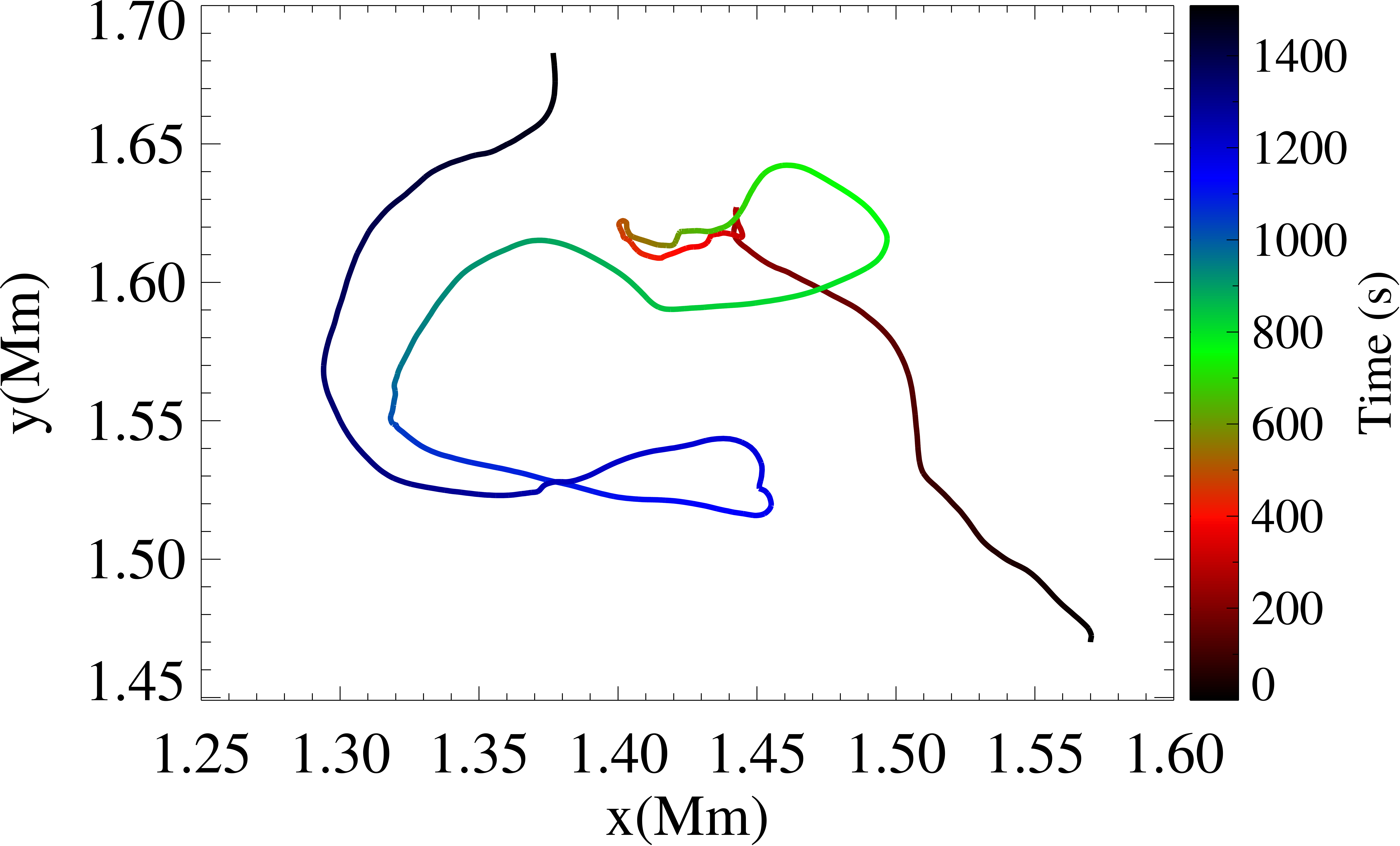}
     \caption{The trajectory of the magnetic field line at the mean solar surface for the  the seed point indicated by a cyan asterisk in Fig. \ref{bz_maps}.}
     \label{field_line_trajectory}
     \end{figure}
\begin{figure*}
     \centering
     \includegraphics[scale=0.97,trim=1cm 1.4cm 0 1.1cm]{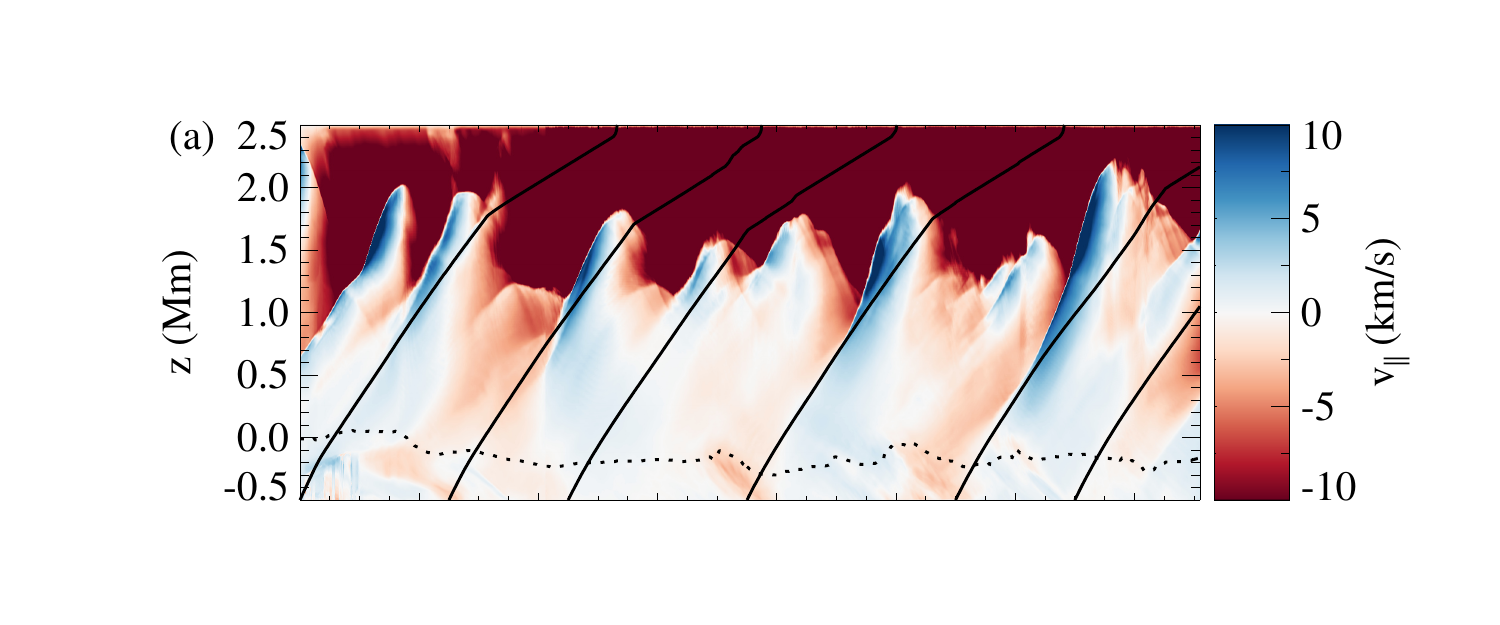}\\
     \includegraphics[scale=0.97,trim=1cm 0 0 1cm]{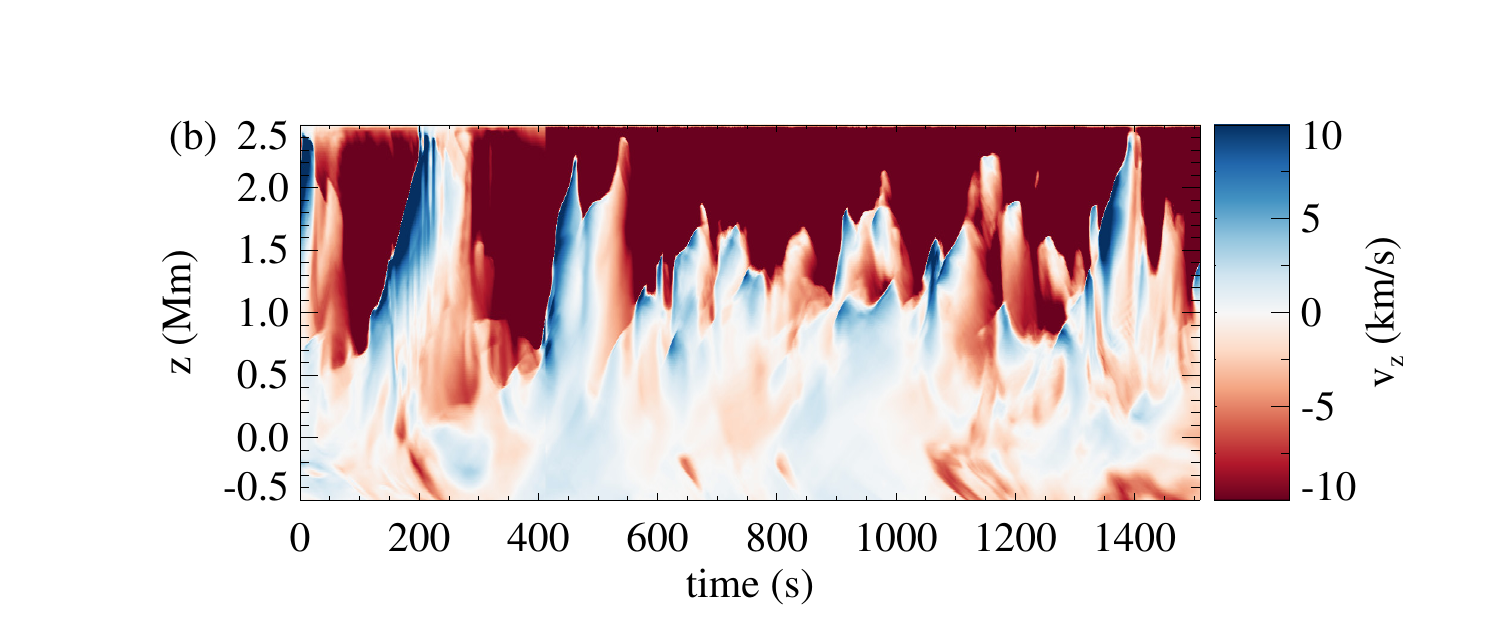}
     \caption{(a): Height-time map of the component of velocity along a magnetic field line that is tracked in time using the method described in sec. \ref{wave analysis}. The field line initially passing through the seed point identified by the cyan asterisk in Fig. 1 is considered here. Local sound speed curves (solid) and a curve showing the height at which $c_s=v_A$ (dotted) are also over-plotted, (b): Height-time map for a fixed horizontal location i.e. initial seed location without following the magnetic field line for the  the seed point shown as cyan asterisk in Fig. \ref{bz_maps}.}
    \label{vs_maps}
\end{figure*}
\begin{figure*}
     \centering
    \includegraphics[scale=0.28]{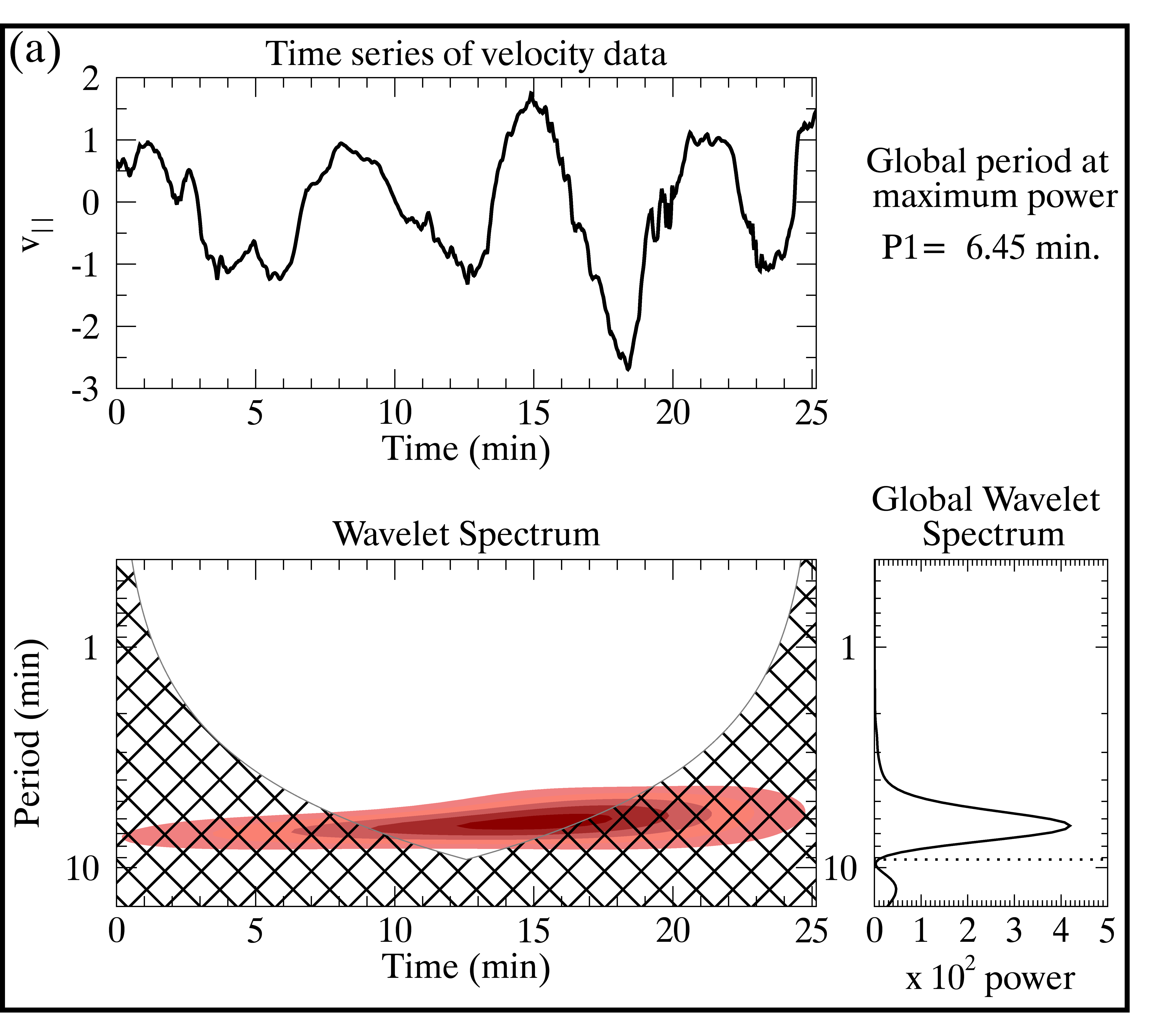}
    \includegraphics[scale=0.28]{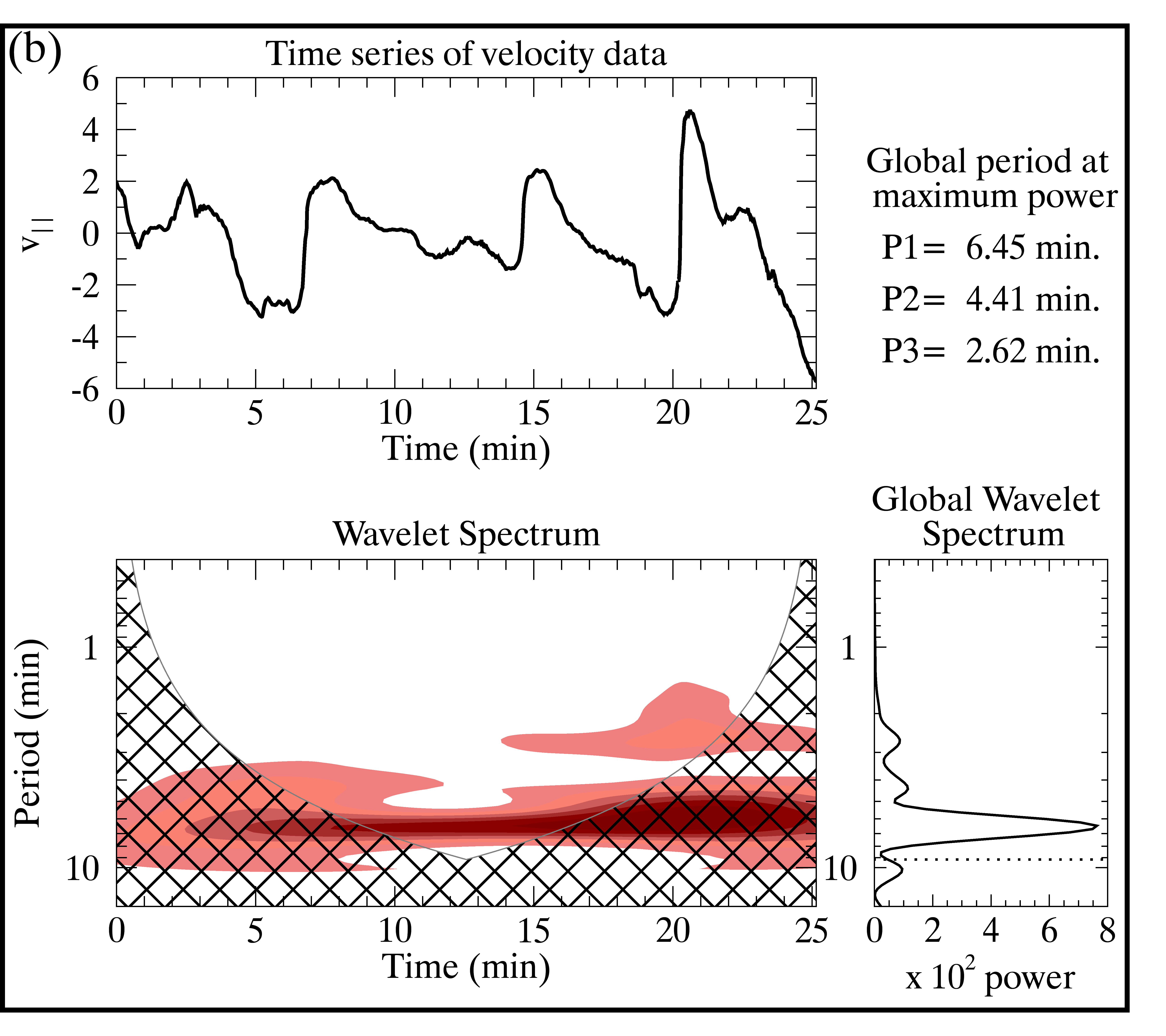}
    \includegraphics[scale=0.28]{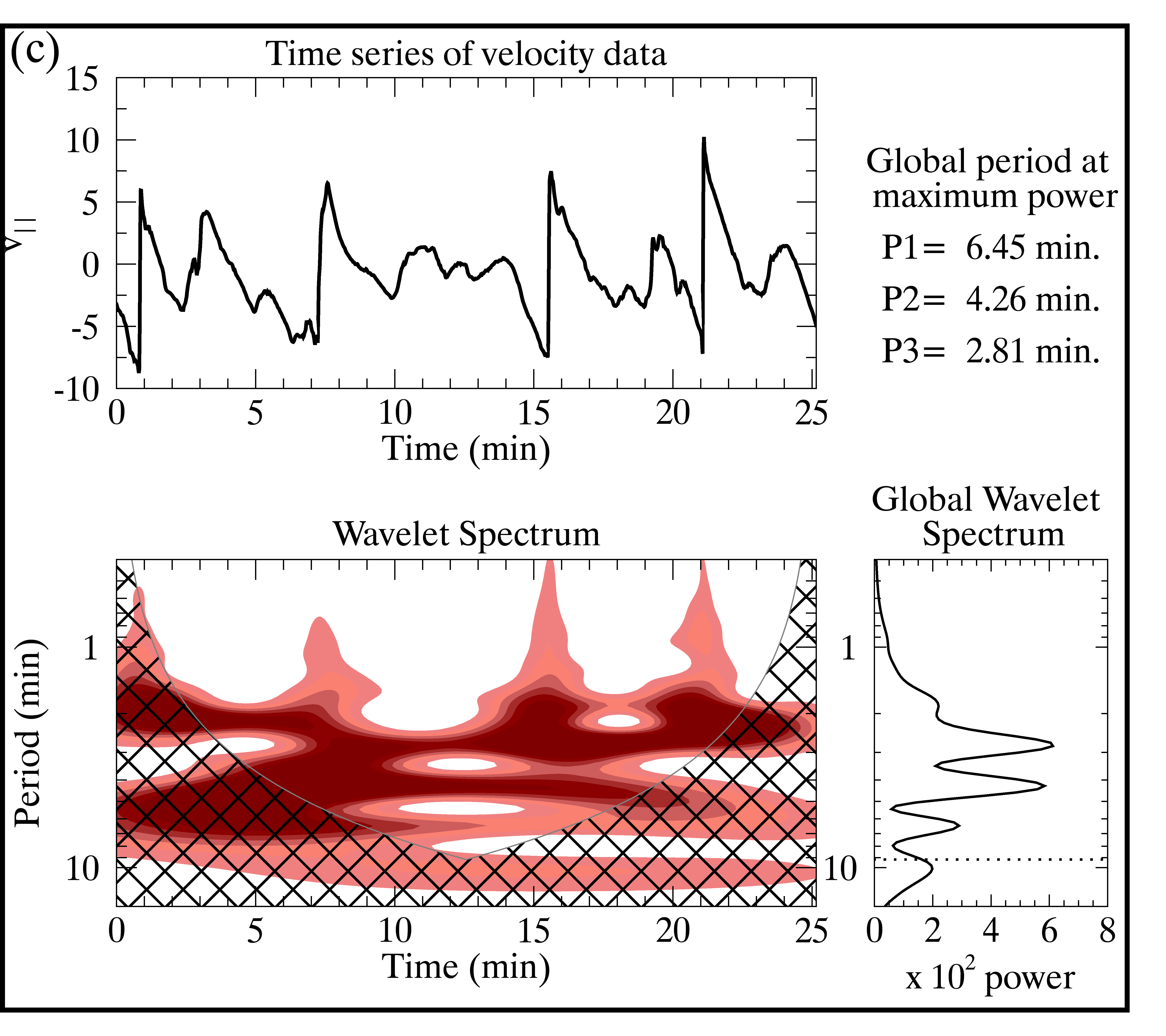}
    \includegraphics[scale=0.28]{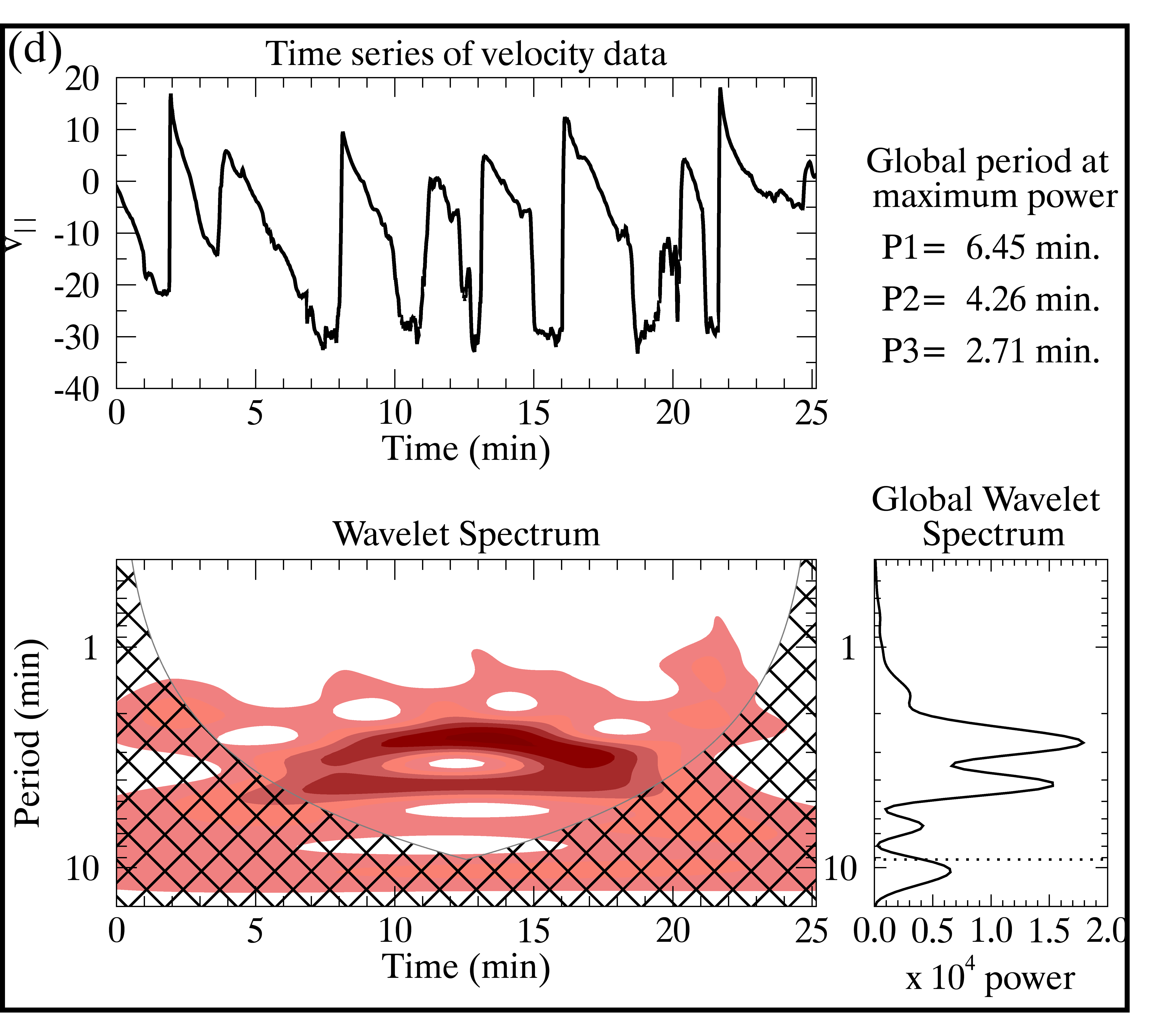}
     \caption{Wave trains (upper sub-panels) and wavelet spectrum of longitudinal (i.e. parallel to the magnetic field) component of the velocity at different heights viz. (a): z=0, (b): z=500 km, (c): z=1000 km and (d): z=1500 km above the mean solar surface for an example case (shown as Cyan asterisk in Fig. \ref{bz_maps}) while tracking the field line in space and time (darker color corresponds to higher power).
     Hatched areas indicate parts of the power spectrum that are less reliable (cone of influence). In the subpanels at the right, global wavelet power spectra are plotted. The horizontal dotted line corresponds to the maximum period for which the power can be determined by the wavelet transform.}
         \label{event25_wavelet}
\end{figure*}
\begin{figure*}
   \centering
    \includegraphics[scale=0.28]{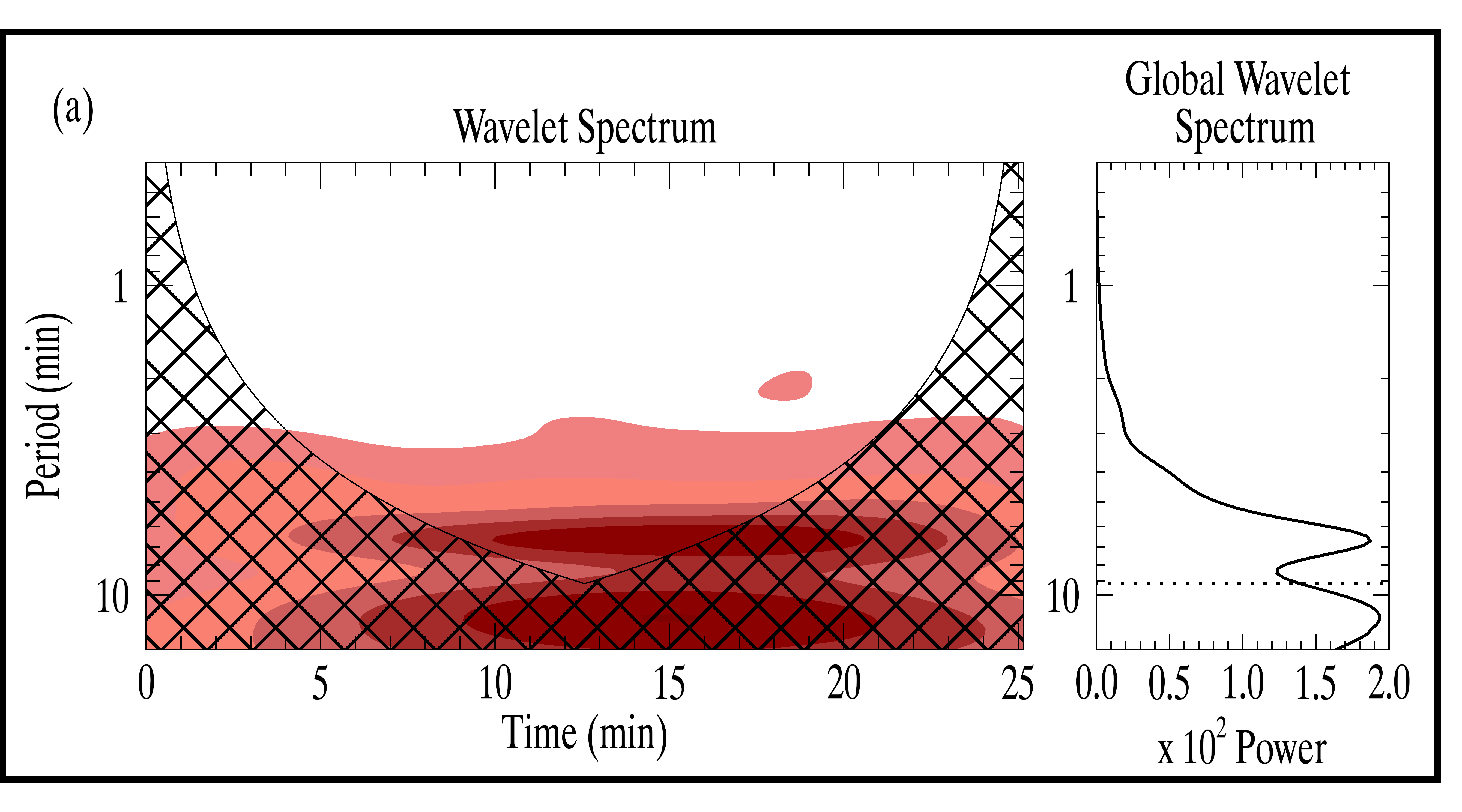}
    \includegraphics[scale=0.28]{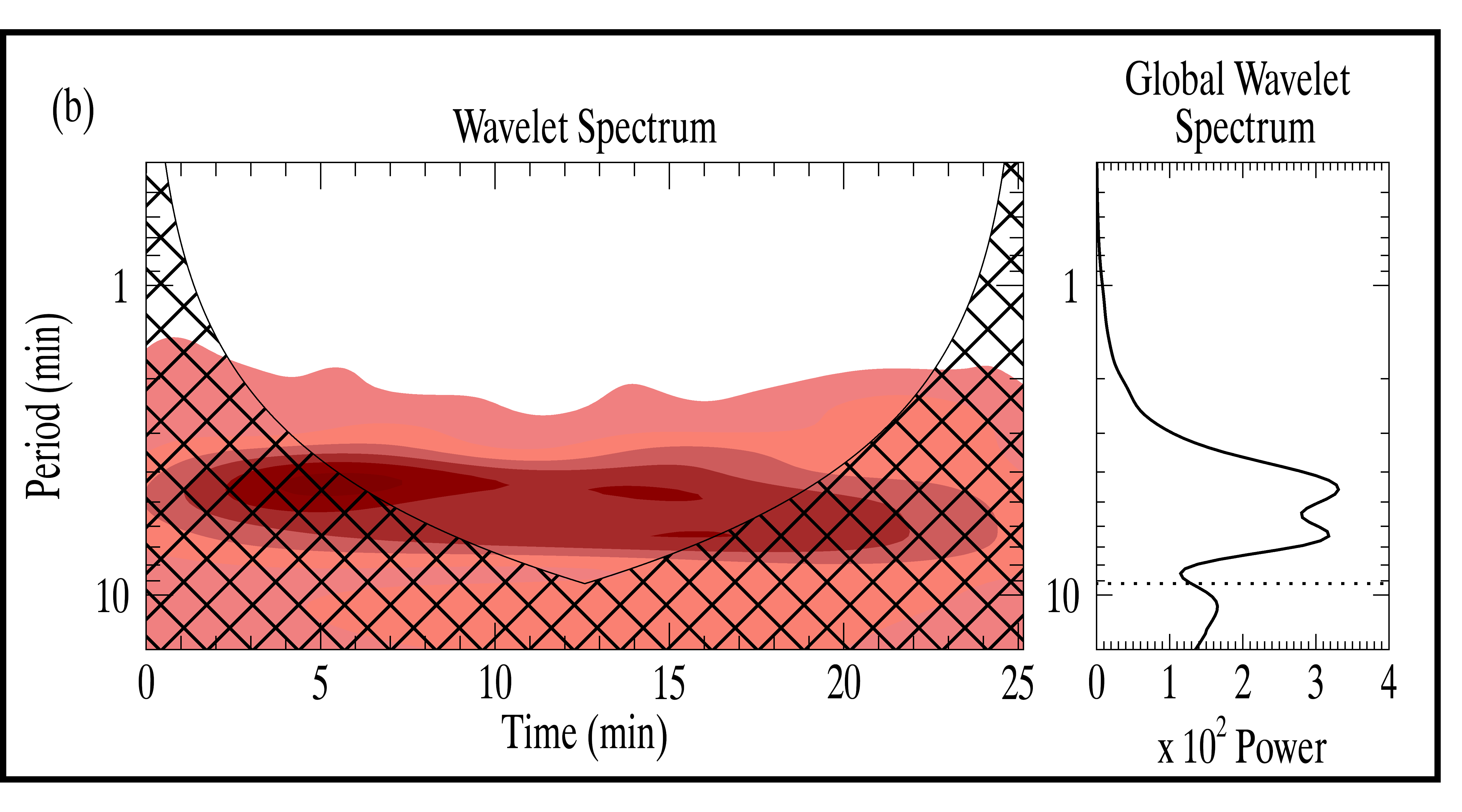}
    \includegraphics[scale=0.28]{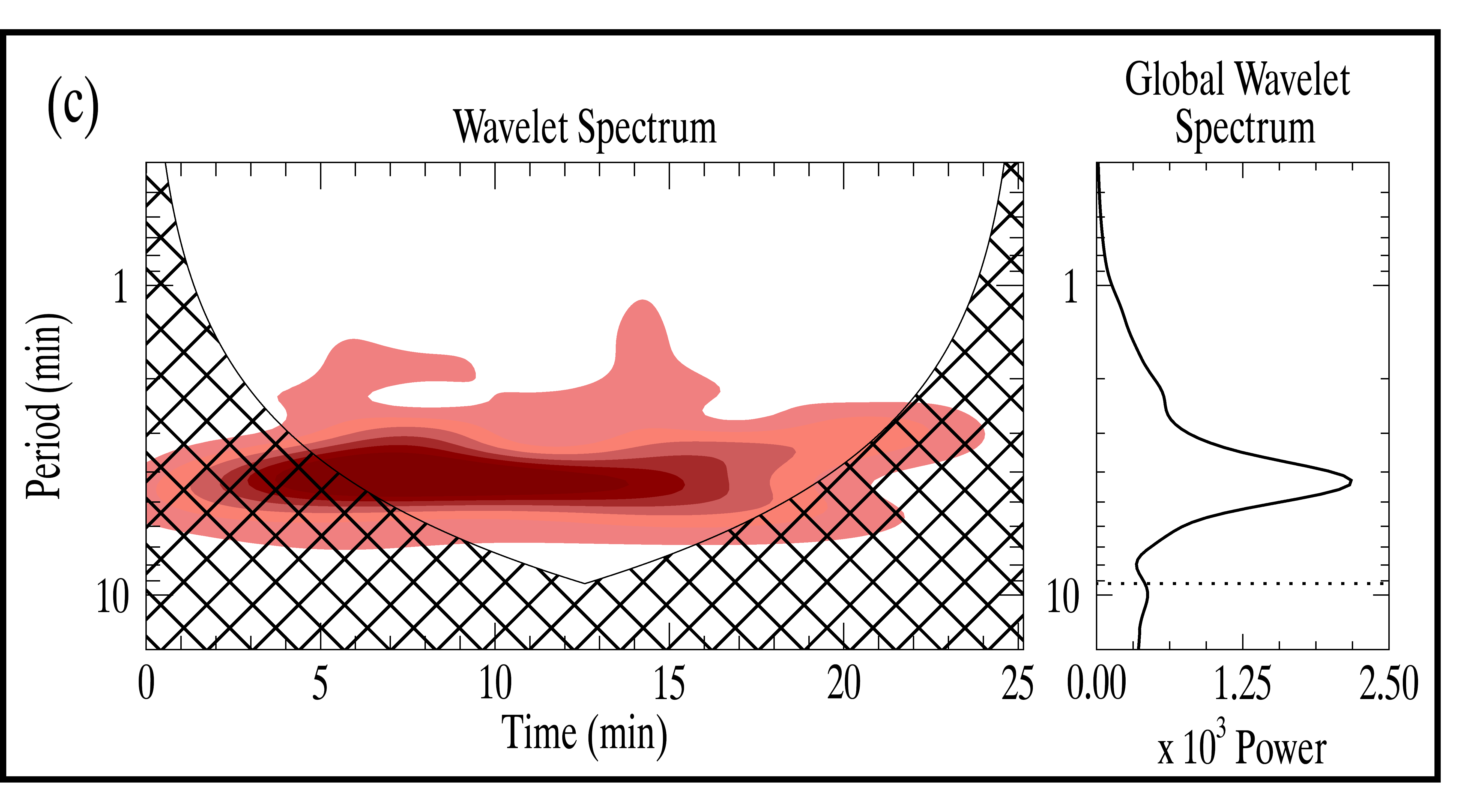}
    \includegraphics[scale=0.28]{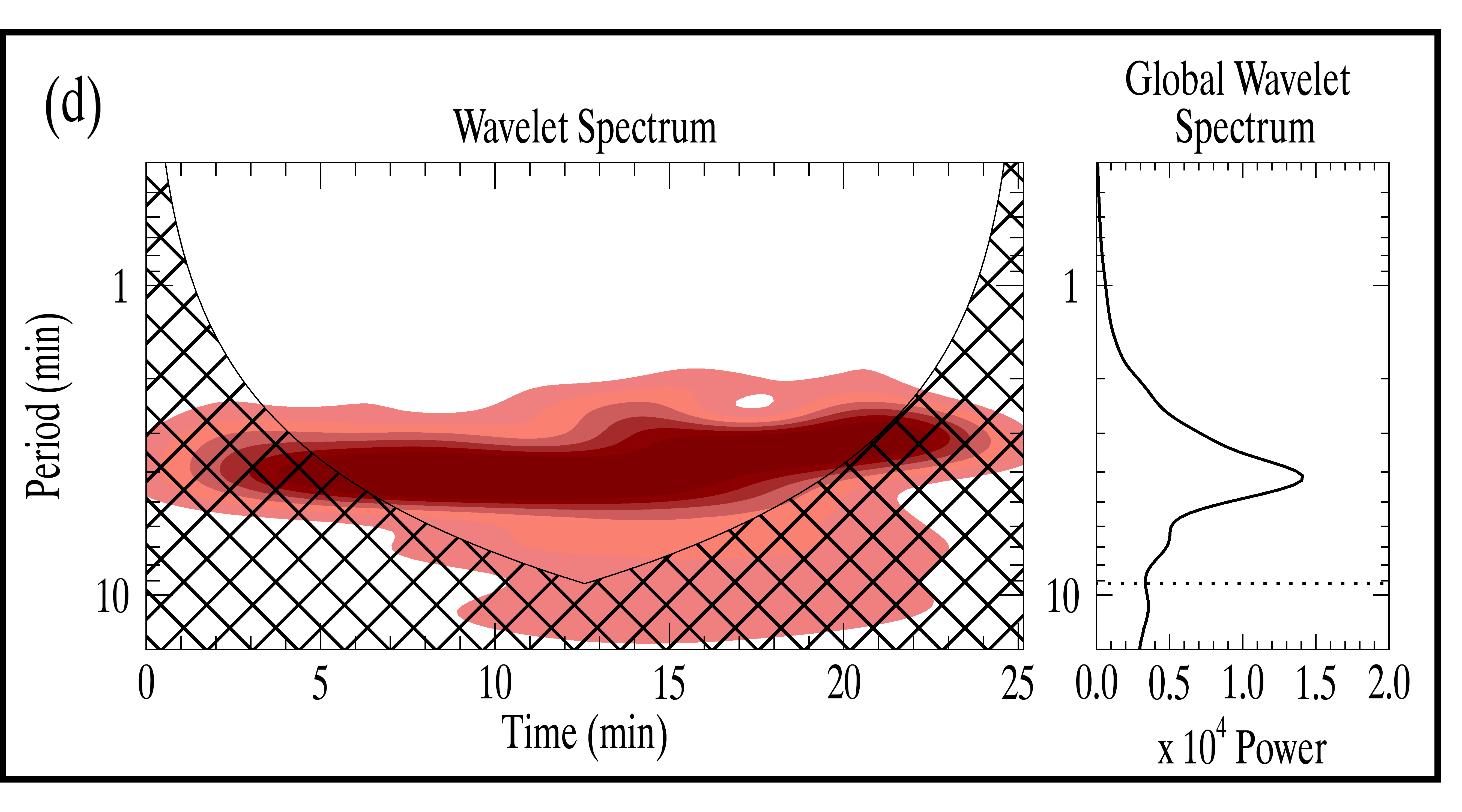}
     \caption{Same as Fig. \ref{event25_wavelet}, but without the panels displaying the wave trains and averaged over all the field lines initially starting from the seed points shown as asterisks in Fig. \ref{bz_maps} (darker color corresponds to higher power)}
         \label{average_wavelet}
\end{figure*}
With this motivation, we performed numerical simulations of a plage region and investigate the propagation of magneto-acoustic waves that are naturally excited due to turbulent convection.
Therefore, unlike in most earlier studies, we do not prescribe the external perturbation of the magnetic features, so that we expect that various wavemodes are excited  at amplitudes similar to those  also present on the real Sun.
These excited slow magneto-acoustic waves travel along the magnetic field lines through the solar atmosphere.
Given that the magnetic field is advected by turbulent granular motions, it is important to discern the temporal evolution of the magnetic field lines for studying waves passing along them. 
We selected a region hosting a wide patch of strong magnetic concentration such that magnetic field extends well into the solar atmosphere.
This allows us to study the waves that could potentially propagate and impact the overlying chromosphere.
To this end, we placed 100 seed points in a magnetically concentrated region at the mean solar surface.
On each seed point, the magnetic field line is traced in 3D space using the magnetic field's vector components at t=0 (the beginning of time-series).
Thereafter, using the velocity vector data at z=0, we calculate the displacement of each seed point (fluid parcel) by integrating in time (here we assume that velocity does not change much in one second).
Then we trace the magnetic field line from the new location of the seed point (which will not necessarily be on the z=0 surface) using magnetic field vector components at the next time step.
Next, we use the intersections of the traced magnetic field lines at t=1 s with the mean solar surface as the seed location for the next time step to follow the field line.
We continue the above procedure and track the magnetic field lines over the entire course of the time-sequence i.e. 25 minutes.
Out of 100 seed points, only 25 (displayed in the right panel of figure \ref{bz_maps}) stayed in the selected domain for the whole time sequence. 
Field lines associated with other seed points were either inclined (such that they extend beyond the selected domain at other heights) or got reconnected due to magnetic reconnection at the edges of the magnetic features or moved out of the domain in the course of 25 minutes and hence were discarded.
A cartoon demonstrating this procedure for one seed point is shown in Fig. \ref{cartoon}.

\section{Results and discussion}\label{3}
As a representative case, we choose the seed point shown in the Cyan color asterisk in Fig. \ref{bz_maps} and display the magnetic field line (green color) attached to it at intervals of 5 minutes in panels (a-f) of Fig. \ref{vapor} (using VAPOR (\citealt{clyne2005}; \citealt{clyne2007}) for 3D visualization).
Saturated maps of the vertical component of the magnetic field strength at the mean solar surface are also displayed for each snapshot.
We find that magnetic field strength at the mean solar surface evolves with time and the magnetic field line undergoes significant displacements from its initial location as its footpoint at the solar surface is advected due to the surrounding plasma motions.

The trajectory of the magnetic field line (corresponding to the selected seed point) where it cuts the $z=0$ plane is displayed in Figure \ref{field_line_trajectory}. 
Here, the color of the curve corresponds to the time elapsed since the beginning of the tracking and the axes show the spatial location.
The magnetic field line travels (due to advection by plasma motions) 1.56 Mm in total with an average horizontal speed of  1.03 km/s over 25 minutes.
Figure \ref{field_line_trajectory} shows that the magnetic field line undergoes several rotations e.g., centered roughly at (1.47,1.62) and (1.42,1.53), indicating torsional motion.
However, in the present paper, we mainly focus on the longitudinal motions of plasma that correspond to slow magneto-acoustic waves. The torsional motions in this simulation have been studied by \citet{yadav2020a,yadav2020b}.

We construct the height-time map of longitudinal velocity to investigate the propagation of slow magneto-acoustic wave in the simulation box.
We calculate the longitudinal velocity at each vertical grid point and track the field line using the method described in sec. \ref{wave analysis}. 
Panel (a) of Fig. \ref{vs_maps} displays the height-time map for one example.
It indicates the propagation of longitudinal velocity perturbations along the magnetic field lines.
Here, the vertical axis is the height normal to the mean solar surface.
Since the selected region under investigation has kG magnetic field strength in the photosphere, the plasma beta is low, and slow magneto-acoustic waves propagate along the field lines at nearly the local sound speed.
To calculate the adiabatic (or isentropic) sound speed i.e. $c_s=\sqrt{{\gamma_1 P}/{\rho}}$, we use local plasma parameters and calculate the adiabatic coefficient ($\gamma_1$) using Chandrasekhar’s first adiabatic exponent (\citealt{chandrasekhar1957}) defined as 
\begin{equation}
\gamma_1=\bigg(\frac{\partial \ln P}{\partial \ln \rho}\bigg)_s
\end{equation}
To compute the adiabatic coefficient we have used the look-up tables for the equation of state (EOS) and method of inversions of the Jacobian matrix (for details, see appendix A.5 in \citealt{cheung2006}).

\begin{figure*}
   \centering
    \includegraphics[scale=0.28]{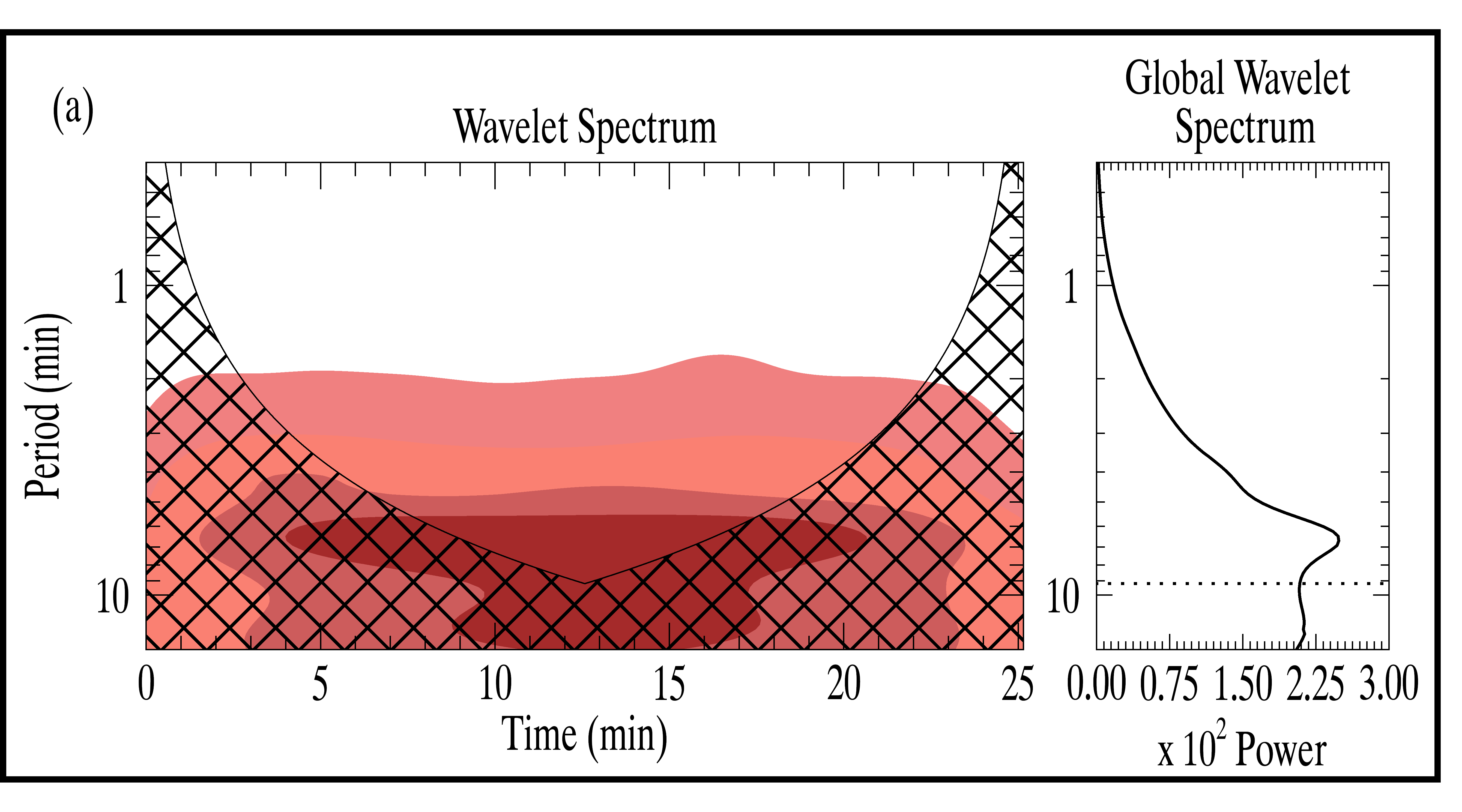}
    \includegraphics[scale=0.28]{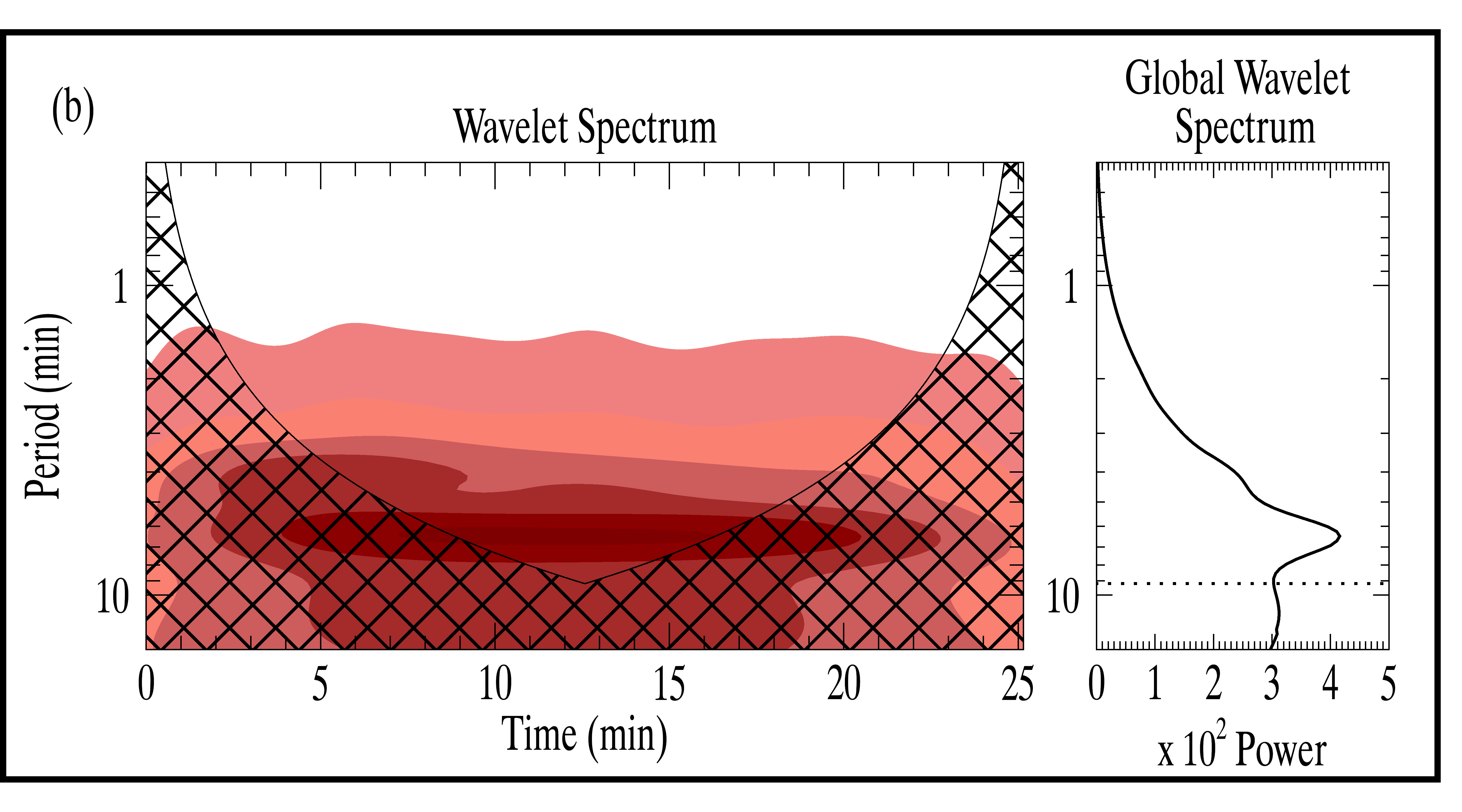}
    \includegraphics[scale=0.28]{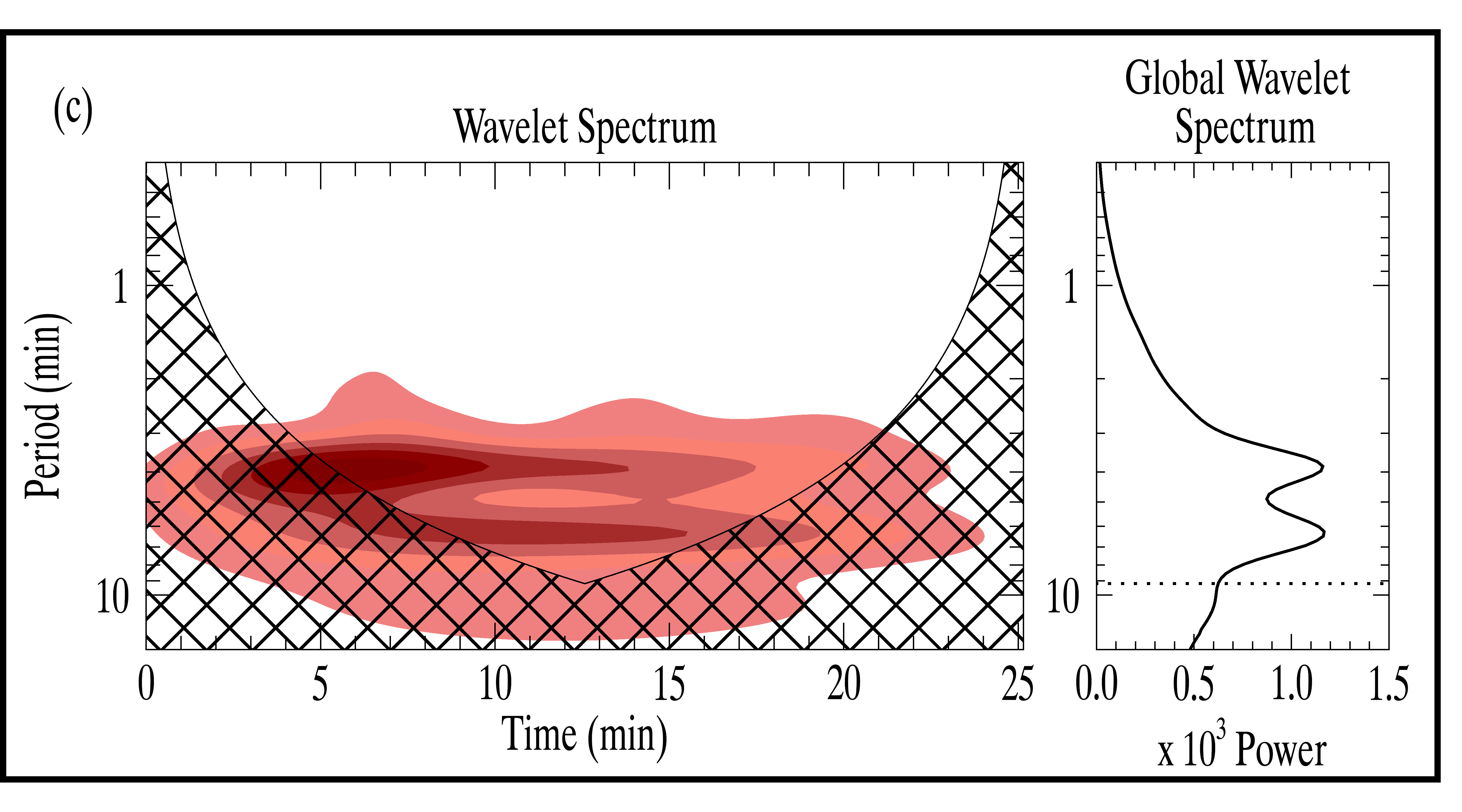}
    \includegraphics[scale=0.28]{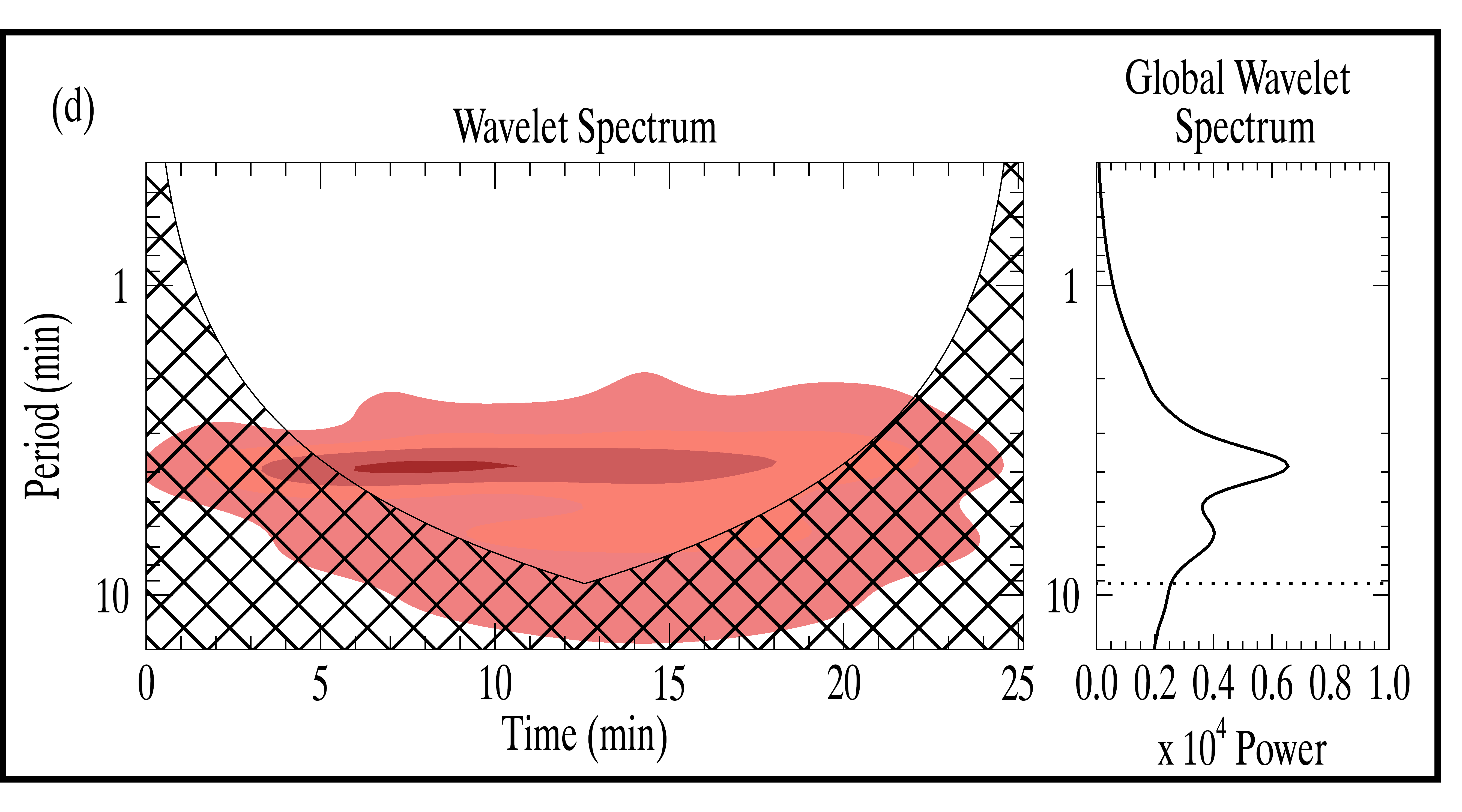}
     \caption{Same as Fig. \ref{average_wavelet}, but averaged over the whole horizontal domain (darker color corresponds to higher power)}
         \label{average_wavelet_whole_domain}
\end{figure*}
For comparison, we over-plot the paths of test particles moving vertically with local sound speed as displayed by solid black lines.
A dotted line is also shown that represents the height where sound speed equals the Alfv\'en speed.
This is called the Alfv\'en-acoustic equipartition layer where helioseismic waves split into fast and slow magneto-acoustic branches (\citealt{cally2007}).
We can see the partial transmission of incident fast acoustic waves into slow magneto-acoustic waves and partial reflection of fast magneto-acoustic waves at this layer.
The slow compressive mode then travels upwards along the magnetic field lines and is converted into large-amplitude shocks due to sharp density gradients in the chromosphere.
These shocks are supersonic and therefore travel faster than the local sound speed as can be seen above a height of $\sim$ 1 Mm.
We find downflows in the layers near the top boundary due to the closed top boundary condition.
Therefore, the results obtained above $\sim$ 1 Mm are influenced by the top boundary condition and hence, not discussed in this paper.
For comparison, we have also displayed a height-time map of the vertical component of the velocity at the initial location of the selected seed point as it would be observed if we did not track the magnetic field line in time (Fig. \ref{vs_maps}b).
We find that the map is not smooth in this case.
The reason for these intermittent fluctuations is that the waves travelling in neighbouring field lines are not in phase (will be discussed later in the paper).
Therefore, oscillations at a fixed location provide a false impression of the high-frequency component.
It justifies and reinforces the importance of tracking the magnetic field lines and of measuring the longitudinal component of velocity to accurately analyze the properties of slow magneto-acoustic waves.
Also, by tracking the magnetic field lines in time, we can determine the locations where linear magneto-acoustic waves become nonlinear and form magneto-acoustic shocks. 

Further, we determine the dominant frequencies of slow magneto-acoustic waves at various heights in the solar atmosphere.
We compute height-time maps and temporal power spectra of longitudinal velocity for all the seed points displayed as white asterisks in the right panel of Figure \ref{bz_maps}.
We use the wavelet transform method as it efficiently captures short-lived features and waves (\citealp{torrence1998,jess2007}).
In a wavelet transform, a time-series of velocity or any other data is convolved with the "mother" function (`Morlet wavelet' in the present case) whose width is varied so that it captures both low and high-frequency oscillations.
The output of the wavelet transform gives the time-period map which reveals the frequency/period distribution of the data at each instant of time.
Averaging over the time axis gives the global wavelet spectrum.
Since we deal with finite time-series data, there will be errors in the beginning and at the end of the time-series.
To avoid it, the time-series is padded with zeros, thus limiting the edge effects.
Zero padding introduces discontinuity near the endpoints.
To select the locations where wavelet power can be fully trusted, a \textit{cone of influence} (COI) is calculated.
The COI defines the region in which edge effects are significant.
Beyond the COI, edge effects are negligible and wavelet power is reliable.
A Fourier transform was also performed for comparison and verification of our wavelet transform results.
However, we have only presented the results obtained with the wavelet transform in the paper as they are more appropriate for wave studies in the solar atmosphere, with typically short wavetrains.

Figure \ref{event25_wavelet} displays the temporal power spectra of longitudinal velocity at four heights viz. z=0, z=500 km, z=1000 km and z=1500 km in panels (a-d), respectively, for the selected representative case.
We have displayed time-series of the longitudinal velocity, time-period map of the wavelet power and time-averaged global wavelet spectrum at each height.
The horizontal dotted line over-plotted on the global wavelet spectrum displays the maximum period for which the power can be determined by the wavelet transform.
At heights of z=0 and z=500 km, the longitudinal velocity has a linear oscillatory profile whereas at z=1000 km and z=1500 km the waveform is no longer sinusoidal, and signs of shocks are clearly visible at these heights.
In the near-surface layers, the amplitude of longitudinal velocity is smaller and the waves have periods around 6-7 minutes.
These high periods of velocity oscillations are possibly due to strong magnetic fields in the selected domain as the period of oscillations increases with the strength of magnetic field in solar plage (\citet{2013Kostik,2016Chelpanov}).
In the higher layers, wave amplitudes increase (due to density stratification), and there is more power in comparatively smaller wave-periods (higher frequencies).
One reason behind the shift of peak frequency could be the frequency filtering by the upper layers that only allow waves with a frequency higher than the local cut-off frequency to pass through (\citealt{1973SMichalitsanos,1977bel}).
It is clear from panel (c) and (d) that shocks are the dominant source for the enhanced power in lower periods or higher frequencies as previously suggested by \citet{2009vecchio}

To estimate the power spectra in a typical strong magnetic element with mostly vertical magnetic field lines, we take an average of frequency spectra calculated at all the seed locations (shown as white asterisks in Fig. \ref{bz_maps}).
The selected 25 field lines are mostly vertical as we select only those field lines which do not bend too strongly and stay fully in the selected domain ($\mathrm{3 Mm \times 3 Mm}$ in the horizontal plane) for the whole time-series. 
The averaged wavelet spectra, calculated by taking  an average of 25 individual power spectra, are displayed in Fig. \ref{average_wavelet}.
At z=0, power distribution peaks at around 6 minutes and this peak merges with another peak at around 10.5 minutes. 
Since our time-series lasts 25 minutes, results with periods beyond 9 minutes are not reliable due to the significant edge effects.
Higher in the atmosphere, the distribution shifts towards lower wave-periods ($\sim$4 minutes).

However, it has been previously shown in observations that low frequency waves can leak to higher layers along the inclined magnetic field lines at the edges of strong magnetic elements in plage (\citealt{centeno2009,2010wijn,2019rajaguru}).
To test this wave leakage in our simulations, we compute power spectra in a similar fashion as done for Fig. \ref{average_wavelet}, however, by taking an average over wavelet power spectra calculated over the whole analyzed horizontal domain (3Mm $\times$ 3Mm).
The resulting wavelet spectra are shown in Fig. \ref{average_wavelet_whole_domain}.
Here, we indeed see that a significant fraction of wavelet power at wave period $\sim$ 6.5 minutes is leaking to the higher atmospheric layers which was absent in the averaged spectra over predominantly vertical field lines (Fig. \ref{average_wavelet}).

\begin{figure}
    \centering
\includegraphics[scale=0.56]{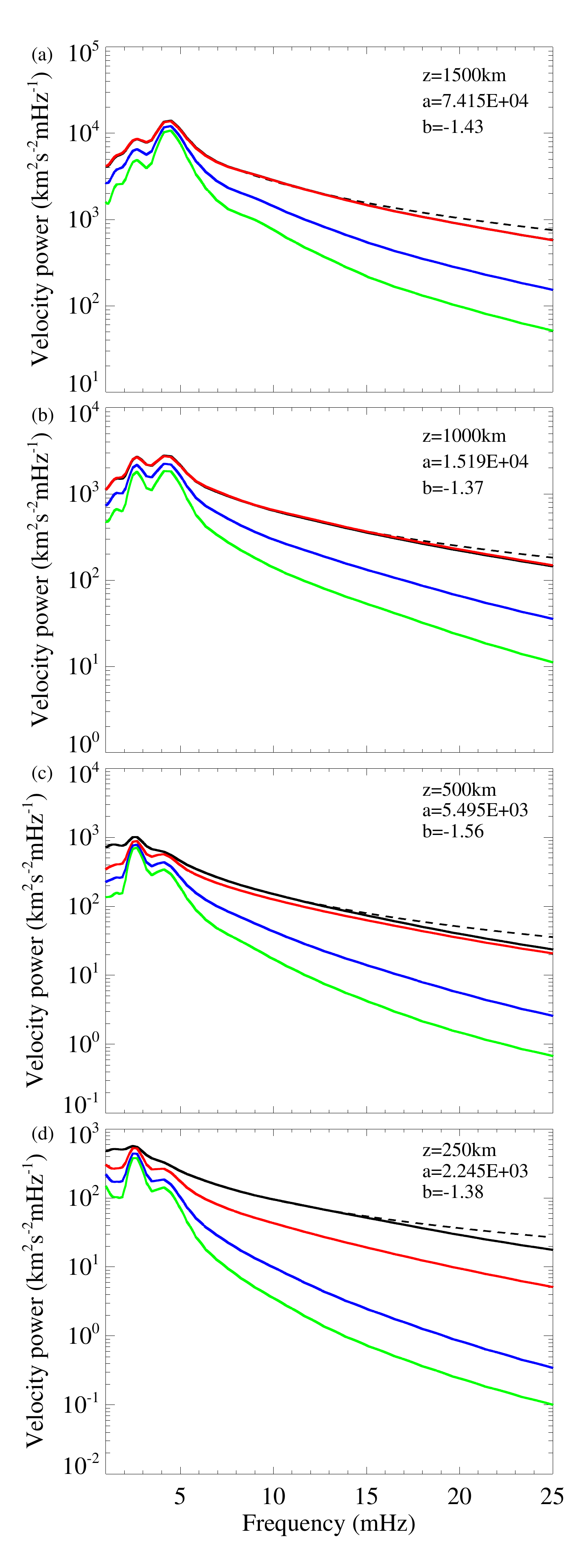}
 \caption{Power spectra of longitudinal velocity, i.e. velocity parallel to the field (black) and vertical velocity (red) at the full resolution of simulations. Power spectra of vertical velocity for degraded simulation data with an effective resolution of 100 km (blue) and 200 km (green). Dashed curve displays a power-law fit to the black curve in the frequency range 6-25 mHz in each panel. Different panels correspond to different heights as mentioned in each panel. ‘a’ and ‘b’ are the amplitude and the exponent for the power law.}
    \label{velo_power} 
\end{figure}
\begin{figure*}
    \centering
    \includegraphics[scale=1.0,trim={0 0 1.5cm 0}]{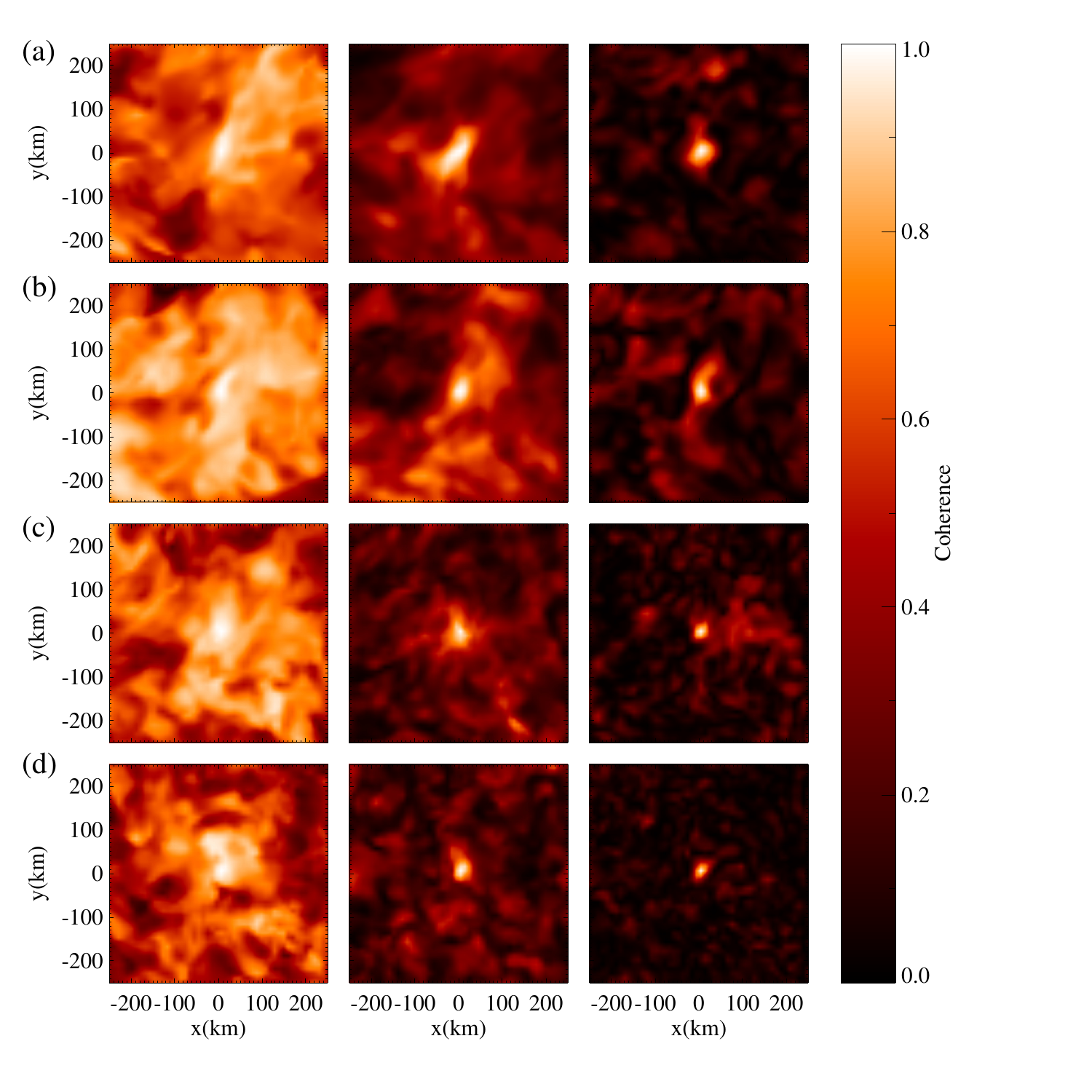}
  \caption{2D maps of coherence between the time-series at the initial seed location (identified by the cyan asterisk in Fig. \ref{bz_maps}) and at surrounding locations at different heights viz. (a) 1500km, (b) 1000km, (c) 500km and (d) 250 km above the mean solar surface.
  The plotted region corresponds to the dashed square in the rightmost panel of Fig. \ref{bz_maps}. 
  Left, middle and right columns correspond to average coherence in the frequency range of 2-5 mHz, 6-10 mHz, 20-25 mHz, respectively.}
    \label{coherence} 
\end{figure*}
Next, to calculate the frequency spectra, we convert the horizontally averaged (over the whole analyzed domain i.e. $\mathrm{3 Mm \times 3 Mm}$)  wavelet spectra to frequency spectra and shown them in Fig. \ref{velo_power}.
Here, black curves correspond to spectra calculated using the longitudinal component (i.e. parallel to the magnetic field) of velocity at fixed locations.
Different rows correspond to different heights above the mean solar surface as mentioned in each panel.
Since it is not yet possible to achieve a spatial resolution of 10 km with observations, we degrade our simulation data to achieve an effective resolution of 100 km and 200 km.
To this end, we convolve the actual simulation data with Gaussian kernels of FWHM of 100 km and 200 km, respectively.
We then calculate horizontally averaged frequency spectra of the vertical component of velocity for full resolution (red) as well for the degraded data-sets at effective resolutions of 100 km (blue) and 200 km (green) as shown in  Fig. \ref{velo_power}.
A power law fit to the black curve is over-plotted as a dashed curve in the frequency range 6-25 mHz in each panel.
We find power spectra to follow a power-law of the form $a f^{-b}$ at all the heights. 
Here, $f$ denotes the frequency, and `$a$' and `$b$' are the amplitude and the exponent for the power law that are specified in each panel.

We find that power spectra have two distinct peaks centred around 2.5 mHz and 4.5 mHz, that correspond to wave periods of $\sim$ 6.5 minutes and $\sim$ 4 minutes, respectively.
Comparing these spectra with the spectra shown in Fig. \ref{average_wavelet}, we notice that there is a significant amount of low frequency waves ($\sim$ 2.5 mHz) leaking up to the higher layers that is absent in Fig. \ref{average_wavelet}. 
This leakage of low frequency waves could be due to expansion of magnetic field lines at the edges of magnetic concentrations.
This discrepancy in wave periods for vertical and non-vertical fields was previously observed and reported by \citet{2010wijn} showing low frequency waves only to propagate at the periphery of the plage.
Moreover, comparing the power spectra of the velocity along the field line and in the vertical direction, we find that the power in the parallel component of the velocity is always higher than in its vertical component in the lower atmospheric layers.
This is expected as slow magneto-acoustic waves travel along the field lines.
However, in the upper layers where magnetic field lines are almost vertical, both red and black curves are nearly identical, implying that the longitudinal component of the velocity can be approximated by the vertical component in these layers.
Here, it is worth mentioning that plasma flows in the atmospheric layers close to the computational domain's top boundary are strongly affected by the imposed closed boundary condition.
Nonetheless, we showed the velocity power spectra in layers at 1 Mm and 1.5 Mm in figures (\ref{event25_wavelet} - \ref{velo_power}) to demonstrate that low frequency waves cease to propagate into higher layers when traveling along the predominantly vertical magnetic field lines, while we find their signatures in power spectra when studying wave propagation over the whole horizontal domain.

Power spectra in the high frequency range (6-25 mHz) approximately obey a power-law and are consistent with the observational findings by \citet{reardon2008} who investigated the acoustic properties of the solar chromosphere in the quiet Sun using observations recorded by the Interferometric Bidimensional Spectrometer (IBIS, \citealt{Cavallini2006}).
They calculated power spectra of the Doppler velocity of the Ca II 854.2 nm line and found that the power spectrum in network regions peaks in the 3-4mHz range (high 5-minute power) and obeys a power-law with a slope of -2.4 in the range of 5–15 mHz.
Although their analysis concentrated on the quiet Sun, their results for network regions can be compared with our simulation results for a plage, as both network and plage host kilogauss magnetic field concentrations and are shown to share similar velocities for equally sized magnetic patches (\citealt{2019buehler}).
They suggested turbulence (generated due to shocks) as the possible source of the chromospheric high-frequency power rather than propagating waves.
This could be the case in our study also as there are magneto-acoustic shocks forming that may generate chromospheric turbulence.
Another possible origin could be the interaction between low-frequency waves that may generate high-frequency waves.
The exact source of the significant power in velocities at higher frequencies is yet not clear.

Moreover, comparing the power spectra of vertical velocity obtained using actual and degraded data, we find that peaks are at similar locations but the power is lower in the degraded data, particularly in the higher frequencies.
To understand the underlying cause of this frequency-dependent behavior, we analyzed coherence between time-series at a fixed location with its neighboring pixels.
The coherence between two time-series is defined as the wavelet cross-spectrum normalized with respect to the wavelet spectrum of individual time series.
It is a measure of the correlation, as a function of frequency, between two time-series of the vertical velocity at two locations.
We find that coherence between two nearby pixels decreases with increasing frequency.
For illustration, in Fig. \ref{coherence}, we show 2D maps of coherence (averaged over a frequency range) between the time-series at the initial seed location (shown by the cyan asterisk in Fig. \ref{bz_maps}) and at its surrounding pixels.
Here rows (a-d) correspond to the same heights for which velocity power spectra were shown in the Fig. \ref{velo_power}. 
Left, middle and right columns correspond to average coherence in the frequency range of 2-5 mHz, 6-10 mHz, 20-25 mHz, respectively.
The selected seed point lies at the center of the maps, where the coherence is unity.
Comparing the three columns we can see that spatial coherence is higher for the lower frequency range relative to the high frequency range.
This implies that high frequency waves get out of phase over smaller horizontal distances compared to low-frequency waves.
This means that observations, even those with a relatively high spatial resolution of 100-200 km, miss a significant amount of power at high frequencies. 
Therefore, high-frequency waves are more affected by the smearing of data as seen in Fig. \ref{velo_power}.
E.g., in the photosphere (at a height of 250 km), the loss in power at the peak frequency ($\sim$2.5 mHz) is only 30 \%\ while at 15 mHz only 5\%\ of the actual power will be observed at a resolution of 100 km.
Also, comparing different rows of Fig. \ref{coherence}, we can see that the spatial extent of coherence increases with height due to expansion of magnetic fields with height.
It explains the comparatively smaller drop due to smearing at high frequencies at greater heights (Fig. \ref{velo_power}).
E.g., at a height of 1500 km, observations with 100 km resolution would lower the power at the peak frequency ($\sim$4.5 mHz) by 20 \%\ while at 15 mHz $\sim$35 \%\ of the actual power will be observed.

  \begin{figure}[t]
  \centering
     \includegraphics[scale=0.32]{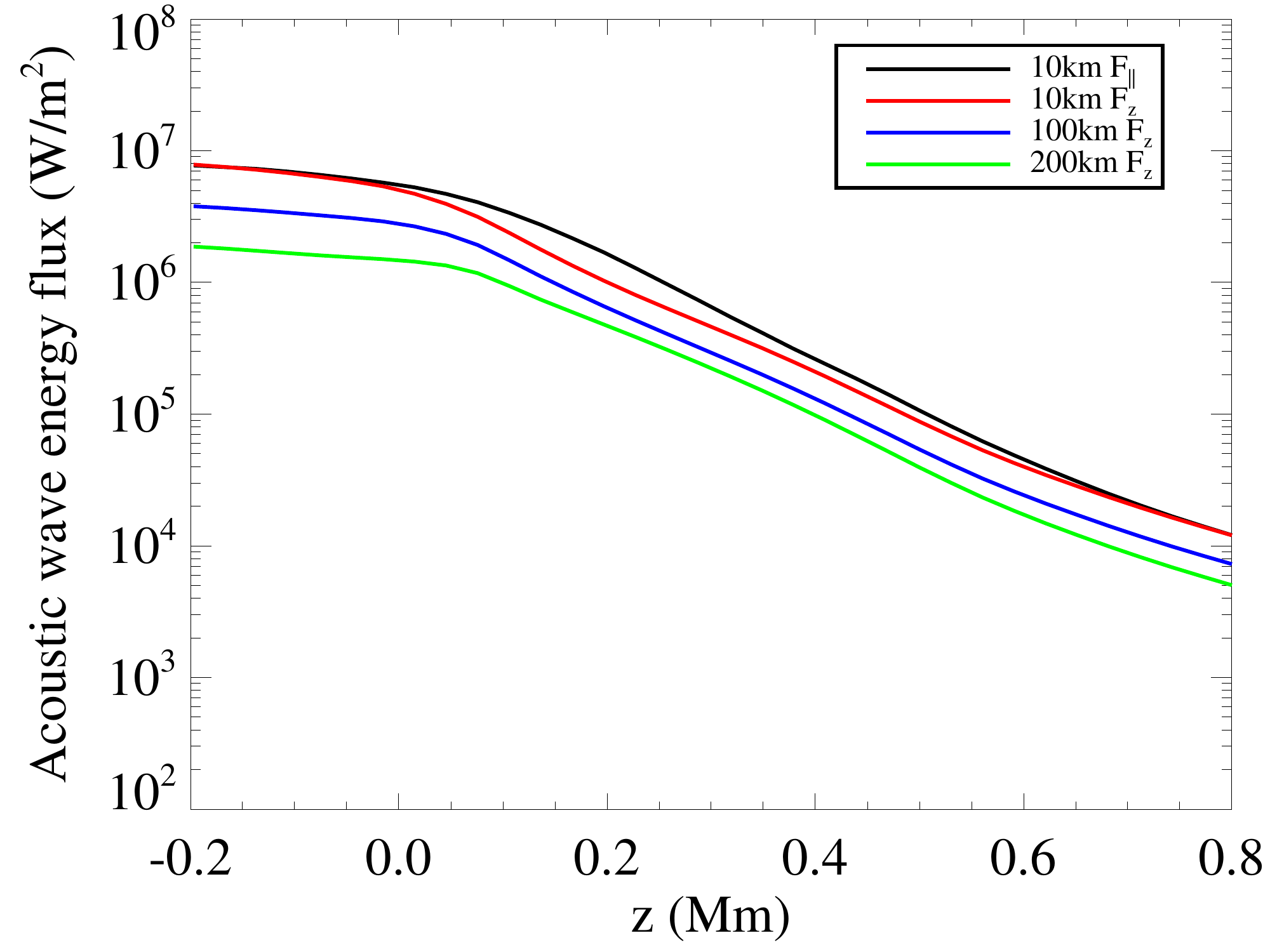}
     \includegraphics[scale=0.32]{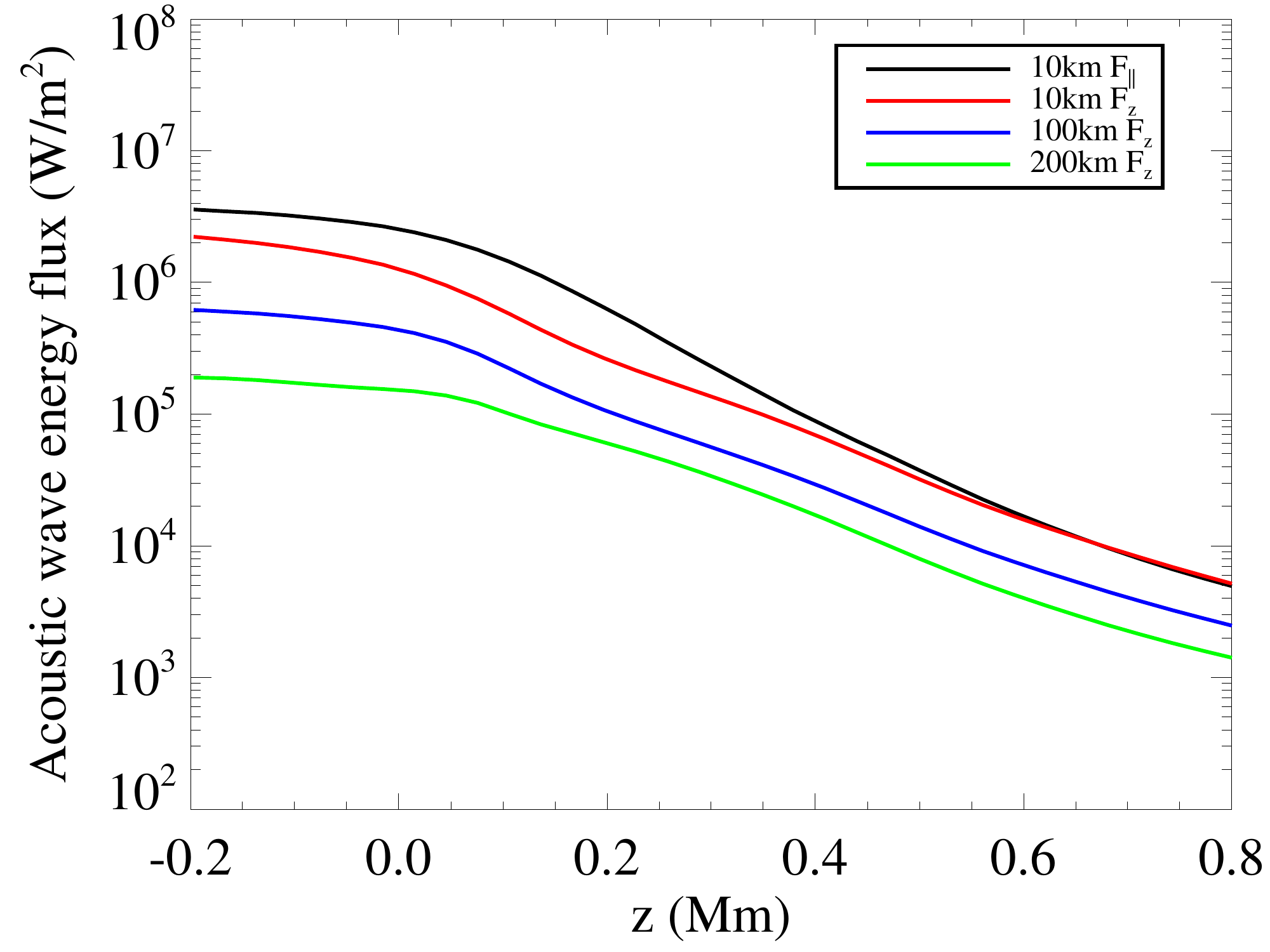}
\caption{Variation of acoustic wave energy flux with geometrical height computed using the component of velocity parallel to the magnetic field for full resolution (black), computed using the vertical component of the velocity for full resolution (red), computed using the vertical component of the velocity for degraded data at an effective resolution of 100 km (blue) and 200 km (green). Left (Right): Integrating velocity power spectra within the frequency range 2-500 mHz (5-500 mHz). }
         \label{wave_energy_comparison}
  \end{figure}
  
To investigate the energy contribution of slow magneto-acoustic waves to the chromosphere, we calculate the acoustic energy flux using velocity power spectra , i.e. $F_{ac, tot}(\nu_i)=\rho \sum_{i}^{} P_{\mathrm{v}} (\nu_i) c_s$ (\citealt{1973Canfield,gonza2009}) and display it as a function of height in Fig. \ref{wave_energy_comparison}.
The height range is shown only up to 800 km for the following reasons.
Plasma down-flows resulting from the artificial closed upper boundary condition can produce non-physical results above this height.
Moreover, the relationship used to calculate the acoustic energy flux is valid only for linear waves and magneto-acoustic waves become nonlinear in nature approximately above this height.
Wave energy flux decreases with height due to the combined effects of wave-reflection, mode-conversion and dissipation.
Since the atmospheric cutoff frequency in facular regions is reduced to $\sim$ 2 mHz (\citealt{centeno2009}), we integrate the velocity power over frequencies larger than 2 mHz for the plage region studied here.
The height variation of acoustic wave energy flux accounting for oscillations with frequencies higher than 2 mHz is shown in the left panel of Fig. \ref{wave_energy_comparison}.
We find higher energy flux when using the parallel component of the velocity compared to the vertical component in the near-surface layers, but the two become very similar in the chromosphere as magnetic field lines are more or less vertical above 500--600 km for unipolar plage with 200 G average field.
Not surprisingly, degrading the spatial resolution reduces the estimated wave energy flux by a factor of 2-3 for the two resolutions tested here. 

The height range from $\sim$ 500 km to $\sim$ 800 km falls in the lower chromosphere as can be seen from Fig. 8(e) of \citet{yadav2020b} which displays the mean temperature stratification for the same simulation setup.
Therefore, the energy flux available at a height of 800 km (i.e. $\sim{10^4 W/m^2}$) is just sufficient to balance the estimated radiative losses in the layers above this height, as estimated by \citet{withbroe1977}.
This is in contrast to previous studies where acoustic flux was found to be insufficient to compensate for the radiative losses (\citealt{fossum2005,2009_beck,Abbasvand_2020,abbasvand2021iris}).
Additionally, the acoustic wave flux in the height range of 250--600 km is much larger than estimated in previous studies pertaining to quiet Sun regions (\citealt{2002_wunnenberg,2008_straus,gonza2009}).
This is due to two reasons: firstly, the high spatial and temporal resolution of our simulations which efficiently capture high-frequency oscillations that are missed in lower resolution observations.
Secondly, facular regions (plage) allow the propagation of low frequency waves along the many inclined field lines near the edges of magnetic elements, so that these waves can reach upper chromosphere. 

To compare our results with the acoustic wave energy flux in the literature, which refers to the quiet Sun, we applied a high-pass filter at the atmospheric cutoff frequency in the quiet Sun of $\sim$ 5 mHz.
The height variation of acoustic wave energy flux accounting solely for oscillations with frequencies more than 5 mHz is shown in the right panel of Fig. \ref{wave_energy_comparison}.
High frequency waves are more influenced by the spatial degradation because they have a smaller spatial coherence, as shown in Fig. \ref{coherence}, and are only partially captured in the spatially degraded data.
Therefore, including the energy contribution only from the high frequency waves shows that degrading the resolution reduces the estimated wave energy flux by a factor of 4-6.
The values for acoustic wave energy flux for degraded data (above 5 mHz) are similar to previously reported values for quiet Sun regions. 
For example, at a height of 600 km above surface, \citet{2002_wunnenberg} estimated an acoustic flux of ${3.6 \times 10^3 W/m^2}$ into the non-magnetic chromosphere. Similar values were reported by \citet{gonza2009} and \citet{2008_straus}. 
We too get similar values at these heights for a degraded resolution of $\sim$ 200 km when including waves with frequencies higher than 5 mHz.
However, in the case of facular regions, the cutoff frequency is $\sim$ 2 mHz and thus allows low frequency waves to propagate through the chromosphere, which contribute significantly to the total acoustic flux (\citealt{1993Lites,centeno2009,2010wijn,2019rajaguru}).

With this numerical experiment, we conclude that in the upper photosphere and the lower chromosphere, the acoustic energy flux is under-estimated in the observations due to poor resolution as well as due to not measuring the longitudinal component of velocity.
For example, at a height of 200 km above the mean surface, the energy flux computed using the vertical component of velocity at full resolution is $ 10^6 W/m^2$, which is a factor of $\sim 2$ higher than the energy flux calculated using degraded data at an effective resolution of 200 km ($ \sim 5 \times 10^5 W/m^2$).
Moreover, the energy flux computed using the component of velocity parallel to the field at this height is $\sim 1.7 \times 10^6 W/m^2$ which is another factor of $\sim 1.7$ higher calculated using the vertical component of velocity at full resolution.
This means that observations at solar disc center with 200 km resolution are likely to catch less than a third of the total acoustic energy flux in the photosphere.
We also find that the acoustic wave energy flux available in the lower and the middle chromosphere is $\sim 10^4 W/m^2$ which is just sufficient to balance the estimated energy requirement in active regions in these layers that is also $\sim 10^4 W/m^2$ (\citealt{vernazza1981,withbroe1977}).

\section{Summary and conclusions}\label{4}
In this paper, we investigated the propagation of slow magneto-acoustic waves along magnetic field lines using high-resolution (10 km), high-cadence (1 s) 3D radiation-MHD simulations of a plage region.
We find that the magnetic field lines carrying slow magneto-acoustic waves are continuously advected by the plasma flows in the photosphere.
Therefore, it is important to follow the individual field lines to study properties of the waves propagating along them.
Height-time maps of longitudinal velocity along the field lines show clear signatures of propagating slow magneto-acoustic waves.
We applied a wavelet transform to compute the frequency spectra at various heights.
The dominant frequency  of slow-mode oscillations largely depends on the magnetic geometry of the region. 
Power spectra of longitudinal velocity averaged over the selected seed points in the core of a magnetic concentration show that the peak frequency shifts towards higher frequencies and there is negligible power found around $\mathrm{3\,mHz}$ (possibly due to acoustic cut-off) as we move to higher atmospheric layers.
These selected seed points, however, are associated with mostly vertical magnetic field lines representative of the central parts of magnetic concentrations.
In contrast, there is significant power at frequencies around $\mathrm{3\,mHz}$ in horizontally averaged power spectra over the whole domain.
This is likely due to leakage of low-frequency waves along the inclined magnetic field lines found at the edges of kG magnetic features, as previously suggested by \citet{2010wijn}.

To estimate the energy contribution from high-frequency (f>6 mHz) waves, we calculated power spectra extending up to 25 mHz.
We find that power spectra in the high-frequency range (6-25 mHz) can be approximated by a power-law of the form $a f^{-b}$ (with exponent $b$ lying around 1.5), suggesting wave-excitation due to turbulence.
This is consistent with previous studies where acoustic energy generation is strongly linked to the Kolmogorov turbulent energy spectrum (\citealt{Musielak_1994,Ulmschneider_1996}).
Integrating velocity power over frequency range 2-50 mHz we find that magneto-acoustic waves carry just sufficient energy to heat the chromosphere in solar plage regions. 
This is in contrast to the claims by \citet{fossumnature2005} and recent observations by \citet{Abbasvand_2020}.

The reason for this discrepancy may have to do with our finding that at frequencies above 5 mHz the waves are not in phase within a single magnetic flux concentration (they get out of phase even within a few 10s of km horizontal distance, i.e. not much more than the resolution of the simulations).
This has multiple important consequences. 
E.g., even high resolution observations that resolve a magnetic feature (but not its sub-structure) would strongly underestimate the wave power. 
Also, if the wave is out of phase on field lines that are close to each other, it implies the presence of strong horizontal gradients of velocity, temperature, etc., especially in the chromosphere where the wave amplitude is large. 
Such gradients could lead to additional dissipation and heating (phase mixing). 

To test the first of these consequences, we additionally degraded our simulation data to effective resolutions of 100 km and 200 km, and compared the total energy flux in the degraded data with simulation data at full resolution.
We find that even at a high spatial resolution of 100 km, observations may strongly underestimate the wave energy flux. 
The observations may also miss some of the flux due to the fact that only the line-of-sight velocity is measured and not the velocity along the magnetic field line, which determines the energy flux carried by the wave. 
Together, these two effects may lead to an underestimate of the energy flux by up to a factor of three in the upper photosphere and the lower chromosphere deduced from observations at spatial resolution of 200 km.
At lower spatial resolutions, even more of the wave energy flux will be missed.
E.g., \citet{fossumnature2005} found that wave energy flux of high-frequency acoustic waves to be too low, by a factor of at least ten, to balance the radiative losses in the solar chromosphere.
This could be due to the coarse spatial sampling ($\mathrm{\sim0.5'' pixel^{-1}}$) of the instrument and large width of the response heights of the observed passbands. 

It is important to note that although the new generation of large telescopes, foremost among them the Daniel K. Inoue Solar Telescope (DKIST), but in future also the European Solar Telescope (EST), hopefully largely overcome the loss in wave energy due to limited spatial resolution of observations, they will likely still miss more than half of the longitudinal wave energy flux in plages due to the fact that only the line-of-sight component of the velocity field is measured. 
In quiet Sun regions we expect observations to miss an even larger fraction of the longitudinal wave energy flux, as magnetic features expand faster and the difference between the line-of-sight velocity and the actual longitudinal velocity of the wave is larger. 
\begin{acknowledgements}
The authors thank L. P. Chitta and D. Przyblski for useful discussions on various aspects of this paper. This project has received funding from the European Research Council (ERC) under the European Union’s Horizon 2020 research and innovation programme (grant agreement No. 695075) and has been supported by the BK21 plus program through the National Research Foundation (NRF) funded by the Ministry of Education of Korea.
\end{acknowledgements}
 \bibliographystyle{aa} 
 \bibliography{paper} 
\end{document}